\documentclass[preprint,floatfix
,aps,prd,superscriptaddress]{revtex4}% Preprint

\usepackage[utf8]{inputenc}
\usepackage{graphicx}

\usepackage{tikz}
\usetikzlibrary{shapes,shadows,arrows}

\usepackage{amssymb,amsmath,mathtools,comment,hyperref}

\usepackage{amsthm}
\theoremstyle{plain}
\newtheorem*{theorem*}{Theorem}

\usepackage{enumerate}
\usepackage{appendix}
\usepackage[utf8]{inputenc}
\usepackage{physics}

\usepackage{array, makecell}
\usepackage{cellspace} %
\usepackage{extarrows}
\usepackage{relsize}

\usepackage{bm}
\usepackage{mathrsfs}
\usepackage{xcolor}
\begin{document}

\title{Spectral Properties of the Symmetry Generators 
\\
of Conformal Quantum Mechanics:
A Path-Integral Approach}

\author{H. E. Camblong}
\affiliation{Department of Physics and Astronomy, University of San Francisco, San Francisco, California 94117-1080, USA}
\author{A. Chakraborty}
\affiliation{Department of Physics, University of Houston, Houston, Texas 77024-5005, USA}
\author{P. Lopez Duque}
\affiliation{Department of Physics, University of Houston, Houston, Texas 77024-5005, USA}

\author{C. R. Ord\'{o}\~{n}ez}
\affiliation{Department of Physics, University of Houston, Houston, Texas 77024-5005, USA}

\date{\today}
\begin{abstract}
A path-integral approach is used to study the spectral properties of the generators of the
SO(2,1) symmetry of conformal quantum mechanics (CQM). 
In particular, we consider the CQM version that corresponds to the weak-coupling regime of the inverse square potential.
We develop a general framework to characterize a generic symmetry generator $G$
 (linear combinations of the Hamiltonian $H$, special conformal operator $K$, and dilation operator $D$),
 from which the path-integral propagators follow, leading to a complete spectral decomposition.
This is done for the three classes of operators: elliptic, parabolic, and hyperbolic.
 We also highlight novel results for the hyperbolic operators, with a continuous spectrum, and their quantum-mechanical interpretation.
The spectral technique developed for the eigensystem of continuous-spectrum operators can be generalized to other operator problems.

\end{abstract}
\maketitle

%-----------------------------------------------------------------------------------------------------------------------
\section{Introduction and Context}
\label{sec:intro}

The path-integral approach, along with associated functional techniques, is a powerful methodology that provides a complete characterization of quantum-mechanical systems.
For ordinary quantum mechanics problems, great advances in finding useful solutions have been made in recent decades---and exhaustive lists of path-integral solutions can be found 
in~\cite{Grosche-PI, Grosche-Steiner_PI-QM, Kleinert-PI}.
In this work,
we use the path-integral approach to derive the eigenvectors and eigenvalues of some of the operators relevant to conformal quantum mechanics.

Conformal quantum mechanics (CQM) has attracted considerable attention following its initial formulation in the 1970s, 
when it was first proposed as an example of a scale invariant theory~\cite{Jackiw:72}, and analyzed in detail by de Alfaro, Fubini, and Furlan (dAFF)~\cite{AFF:76} as a (0+1)-dimensional form of conformal field theory. The dAFF model corresponds to the $D=1$ limit of the spacetime $D$-dimensional conformally invariant Lagrangian
density~\cite{AFF:76}
\begin{equation}
{\mathcal L}=\frac{ 1}{2} \partial_{\mu} \phi \partial^{\mu} \phi - g \phi^{2D/(D-2)}
\; .
\label{eq:CFT-general-Lagrangian}
\end{equation}
In the transition to $D=1$, one can interpret the resulting theory as standard quantum mechanics, with a ``field'' $Q(t)$ described as a configuration or position variable subject to an inverse square potential, with Lagrangian
\begin{equation}
L=\frac{ 1}{2} \,  \dot{Q}^2 - \frac{g}{2 Q^2}
\; ,
\label{eq:CQM-ISP-Lagrangian_1d}
\end{equation}
where the dot represents the usual time derivative. 
This original form of CQM, centered on its SO(2,1) conformal symmetry structure, was subsequently used in seminal papers by Jackiw~\cite{Jackiw:80, Jackiw:90}, also including an analysis of a related CQM system of contact interactions~\cite{Jackiw:delta_91}. Systems exhibiting this kind of SO(2,1) conformal symmetry have seen renewed interest in a broad range of physical applications over the years. This is primarily due to its simplicity as a model of conformal field theory and its remarkably wide range of applicability to established physical problems that exhibit approximate conformal symmetry in a window of physical scales.
While not included in the original dAFF model formulation, outstanding realizations of this type of 
inverse-square-potential systems have been found where this symmetry generates a quantum anomaly. A list of noteworthy applications of such systems includes molecular physics~\cite{Cam:mol-anomaly},
black hole thermodynamics~\cite{Cam:BHT,Cam:BHT-SC,Cam:BHT-CT} and 
acceleration radiation~\cite{HBAR-CQM-1, HBAR-CQM-2, HBAR-QO-1, HBAR-QO-2},
the Efimov effect~\cite{Efimov-1, Efimov-2, Efimov_Naidon-Endo-2017, Cam:CQM-ren}, and graphene~\cite{Gorsky:graphene, Ovdat:graphene}, among several others~\cite{Cam:CQM-ren}. 
It is noteworthy that, in addition to the quantum symmetry breaking 
based on a strong-coupling version of Lagrangian~(\ref{eq:CQM-ISP-Lagrangian_1d}), 
another class of quantum anomalies have been found in systems with SO(2,1) conformal symmetry with
contact interactions---they involve 1D three-body interactions in 1D and 2D two-body interactions in 
Fermi systems of ultracold atoms~\cite{Daza-CRO_2D-anomaly,Drut-CRO_1D-anomaly,Maki-CRO_1D-anomaly,delta-prime_2021,Tajima_Cooper-triplets}.

The previous list emphasizes those cases where the CQM interaction is attractive and sufficiently strong, where it leads to a quantum anomaly~\cite{Cam:mol-anomaly,CQM-anom-alg1,CQM-anom-alg2,CQM-anom-alg3}
or some kind of 
renormalization~\cite{Gupta:ISP-ren, Cam_ISP-ren, Cam_DT1, Cam_DT2, Beane:sing-ren, conformality-lost},
or to the fall-to-the-center phenomenon~\cite{Landau:77, MC-SQM}---this regime is called ``strong coupling'' for short.
On the other hand, the form of CQM often discussed in the context of conformal symmetry analyses is based on the dAFF model~\cite{AFF:76}. 
The latter strictly applies to the case when the conformal potential is repulsive (or sufficiently weak, even if attractive) to avert the pathologies inherent for strong coupling $g$ in Eq.~(\ref{eq:CQM-ISP-Lagrangian_1d}), and the conformal symmetry is maintained. 
For the sake of simplicity, in this work we solely focus on this version of CQM (in a slightly generalized format), 
for which we analyze the spectral properties of some of its symmetry generators.
We will address elsewhere the more general case, including the presence of anomalies.

Of particular interest for the present work is the fact that the dAFF model (in the original weak-coupling formulation of the CQM symmetry generators) has been recently found to be a CFT$_{1}$ realization of the AdS/CFT correspondence~\cite{Jackiw_CFT1-1,Jackiw_CFT1-2}, leading to renewed interest in this topic~\cite{CQM_Okazaki-2015,CQM_Okazaki-2017,Pinzul_CFT1-2017,Khodaee_CFT1-2017,deAlmeida_CFT1-2019,CQM-CFT1_Ardon-2021}. 
In addition, using the operators we discuss in this paper, the dynamic evolution within the dAFF model
has been used to study causal diamonds in Minkowski spacetime~\cite{Arzano-1,Arzano-2}. 
The associated physics of finite-lifetime observers, often called diamond observers, 
was first addressed in Ref.~\cite{martinetti-1}
 in a remarkable finding that generalizes a similar thermalization of the vacuum for accelerated observers 
 (the celebrated Unruh effect~\cite{Unruh:76}). Accordingly, these observers have access to only a limited region of spacetime, known as the conformal (causal) diamond; and the vacuum they perceive has a diamond temperature inversely proportional to their lifetime. This insightful result on the thermalization of the vacuum of diamond observers has been further confirmed in Refs.~\cite{martinetti-2, su-ralph-1, su-ralph-2, light-cone, jacobson,Houston_OQS-diamond}.
As there are a number of parallels with the thermodynamic behavior of black hole horizons, where CQM has been successfully used, it is not surprising that that the dAFF model may also be relevant for causal diamonds. Specifically, in Refs.~\cite{Arzano-1, Arzano-2}
the dynamic evolution within the dAFF model was shown to be in correspondence with 
the time evolution of Minkowski observers with a finite lifetime, which is described by 
the radial conformal Killing fields (RCKF) previously developed in Ref.~\cite{RCKF}.
This problem is of interest in further characterizing the causal structure of spacetime, 
and CQM appears to be a promising tool for this purpose. 

\subsection{Scope and Main Results of This Paper}

In the physical applications mentioned above, and especially in regards to the recent work on CQM as 
CFT$_{1}$ and the physics of causal diamonds, the CQM generators play a central role.
For our purposes,
the prototypical CQM generators of generalized time evolution can be taken as the operators $H$, $R$, and $S$, where $H$ is the quantum-mechanical Hamiltonian associated with the 
Lagrangian~(\ref{eq:CQM-ISP-Lagrangian_1d}), and 
$R$ and $S$ are the linear combinations of $H$ and the special conformal generator $K$ involved in 
setting up the Cartan-Weyl basis. 
In particular, the spectral decomposition of the operator $S$, which 
plays a significant role in causal diamonds~\cite{Arzano-1, Arzano-2,RCKF}, is a novel result of our paper.
More generally, the dynamical evolution can be described by any of  
the generalized generators $G$ defined as linear combinations 
of $H$, $D$, and $K$, where $D$ is the dilation operator.

While the usual operator properties of the set $\{H,R,S\}$ and the generalized generators $G$ have been considered 
in the literature~\cite{AFF:76}, a path-integral treatment is lacking.
 It is the purpose of this paper to develop such functional integral approach,
with which we compute the propagators and the spectral properties of the generators $\{H,R,S\}$,
as well as those of the linear combinations that define a generalized generator $G$, falling
 under the three possible classes: elliptic ($R$-like), parabolic ($H$-like), and hyperbolic ($S$-like).

The main results of the paper are summarized below in a convenient format
 that can help identify the key ideas within the extensive mathematical properties being discussed.
\begin{itemize}
    \item
    We have developed a complete Hamiltonian framework for all the generalized generators $G$ of CQM (which proves essential for the path-integral approach).
    \item 
    We have derived the spectral properties, including the eigenvalues and eigenstates for the three classes of CQM generators $G$, viz., elliptic ($R$), parabolic ($H$), and hyperbolic ($S$), using a path-integral approach. This involves
    using different limits of the propagator for the generalized radial harmonic oscillator $K_{l+\nu}^{(RHO)}$ given in Eq.~(\ref{eq:propagator_RHO}).    
     Depending on the classes of the operators, we have used different methods to extract the spectral properties of the operator $G$ from the propagator. 
     Table~\ref{tab:key-equations}
    acts as a pointer to the equations describing the spectral properties found in this paper.
\end{itemize}
\begin{table}[h]
    \centering\setcellgapes{4pt}\makegapedcells \renewcommand\theadfont{\normalsize\bfseries}
    \caption{References to the key results about the spectral properties of the CQM generators.}
    \begin{tabular}{|c|c|c|c|}
        \hline CQM Generator & Propagator & Eigenvalues & Eigenstates \\
       \hline \hline Elliptic ($R$) & $K_{l+\nu}^{(RHO)}$ -- Eqs.~(\ref{eq:propagator_RHO}), (\ref{eq:propagator_ell=RHO}) 
       & Discrete -- Eq.~(\ref{eq:tilde-H_eigenvalues}) & 
       Eq.~(\ref{eq:tilde-H_eigenstates})    
       \\
              \hline Parabolic ($H$) & $\lim_{\omega\rightarrow 0}K_{l+\nu}^{(RHO)}$ -- Eq.~(\ref{eq:propagator_ISP})
   & Continuous in $\mathbb{R}^+$ & Eq.~(\ref{eq:ISP_energy-eigenfunctions}) \\   
       \hline Hyperbolic (S) & $\lim_{\omega\rightarrow -i\omega}K_{l+\nu}^{(RHO)}$ -- 
       Eq.~(\ref{eq:propagator_inverted-RHO})
       & Continuous in $\mathbb{R}$ & Eqs.~(\ref{eq:inverted-RHO_energy-eigenfunctions-1})
       \& (\ref{eq:inverted-RHO_energy-eigenfunctions-2}) \\
              \hline
    \end{tabular}
    \label{tab:key-equations}
\end{table}
\begin{itemize}
    \item We have used a novel method based on Fourier transforms (which we call the Fourier Method) to find the eigenstates of the noncompact CQM operators from the propagator. The key equation for this method is succinctly expressed in 
    Eq.~(\ref{eq:eigenfunctions_from-Fourier_ISP}). Furthermore, we have explored the connection of this method with the retarded and advanced Green's functions in Appendix~\ref{app:Fourier},
 culminating in Eq.~(\ref{eq:eigenfunctions_from-Green}).
 \item
 In particular, we have derived a complete spectral characterization of the operator $S$ and all hyperbolic generators---a result that
 has been surprisingly lacking in the literature.
    \item 
    As a bonus, our detailed analysis of the spectral properties of the propagators via different approaches
    has uncovered additional mathematical connections and identities, as 
    mentioned at the end of Subsec.~\ref{subsec:spectral_H} and in Appendix~\ref{app:Bessel-Whittaker-products}.
    \end{itemize}

\subsection{Organization of This Paper}
\label{subsec:paper_organization}

In Section~\ref{sec:symmetries}, we summarize the symmetry properties of CQM and the specifics of the dAFF model;
 we define the operators $R$, $H$, and $S$ that generate the conformal symmetry group, 
as well as the generic conformal generator $G$; and we develop
the framework that classifies the possible types of $G$ and their role in the dynamical time evolution via their Hamiltonian 
representation.
In Sec.~\ref{sec:PI-setup_CQM}, 
we review the path-integral approach for radial problems in quantum mechanics and specifically use it to study a 
generalized radial harmonic oscillator relevant to CQM.
In Sec.~\ref{sec:spectral_CQM_elliptic},
we fully characterize the spectral properties of the operator $R$ and its
class (elliptic).
In Sec.~\ref{sec:spectral_H_Fourier},
we analyze the spectral properties of the operator $H$ and its
class (parabolic); and we
develop a novel procedure for spectral properties of operators with continuous spectra.
 In Sec.~\ref{sec:spectral_CQM_hyperbolic}, we apply this method to the
conformal operator $S$ and its class of non-compact hyperbolic generators to fully characterize their spectral data and Green's functions.
We conclude the paper in Sec.~\ref{sec:conclusions} with a brief summary and directions for future work.
The appendices cover an alternative dimensional framework for the Hamiltonian description of generators;
detailed properties of the relevant path-integral framework; evaluation of the
inversion integrals for the spectral decomposition of parabolic and hyperbolic generators; 
and further analysis of the conformal generators within a differential-equation approach.
\tikzstyle{decision} = [diamond, draw, fill=blue!50]
\tikzstyle{arrow} = [draw, -stealth, thick] 
\tikzstyle{thick-arrow} = [draw, -stealth, line width=0.5mm]
\tikzstyle{elli}=[draw, ellipse, fill=red!50,minimum height=8mm, text width=5em, text centered]
\tikzstyle{block} = [draw, rectangle, fill=blue!8.5, text width=9em, text centered, minimum height=15mm, node distance=10em]

\begin{figure}
\begin{tikzpicture}
\node [block] (Intro) {Sec.~\ref{sec:intro}:  \\ Introduction \& Context};
\node[block, below of=Intro, yshift=2em](Hamiltonian){Sec.~\ref{sec:symmetries}: \\ CQM Generators \\ \& Hamiltonian Formulation};
\node [block, left of=Hamiltonian, xshift=-1.5em] (App A) {Appendix A: \\ Alternative Hamiltonian Formulation};
\node [block, right of=Hamiltonian, xshift=1.5em] (PI) {Sec.~\ref{sec:PI-setup_CQM}:  \\ Path-Integral (PI) \\ Framework for CQM};
\node[block, below of=PI, xshift=-7.5em, yshift=-1.75em](CQM Gen) {Secs.~\ref{sec:spectral_CQM_elliptic}, \ref{sec:spectral_H_Fourier}, \ref{sec:spectral_CQM_hyperbolic}: \\  PI Spectral \\ Analysis of Generic CQM Generators};
\node [block, right of=PI, xshift=1.5em] (App B) {Appendix B: PI Framework};
\node [block, below of=PI, xshift=4.25em, yshift=1.25em] (App C) {Appendix C: \\ Fourier \\ Method};
\node [block, right of=CQM Gen, xshift=1.75em,yshift=-3em] (App D) {Appendix D: \\ Integral \\ Representations};

\node [block, below of=App A, xshift=2.5em, yshift=-1.75em] (App E) {Appendix E: Differential-Equation Approach};

\node[block, below of=CQM Gen, xshift=-0.05em, yshift=1.5em](Conclusions){Sec.~\ref{sec:conclusions}: Conclusions};

\path [thick-arrow] (Intro) -- (Hamiltonian);
\path [thick-arrow] (Hamiltonian) -- (PI);
\path [thick-arrow] (App B) -- (PI);
\path [thick-arrow] (Hamiltonian) -- (CQM Gen);
\path [thick-arrow] (Hamiltonian) -- (App A);
\path [thick-arrow] (PI) -- (CQM Gen);

\path [thick-arrow] (App C) -- (CQM Gen);
\path [thick-arrow] (App D) -- (CQM Gen);
\path [thick-arrow] (Hamiltonian) -- (App E);
\path [thick-arrow] (App E) -- (CQM Gen);
\path [thick-arrow] (CQM Gen) -- (App E);
\path [thick-arrow] (CQM Gen) -- (Conclusions);

\end{tikzpicture}
\caption{Flowchart of this paper.}
\label{fig:flowchart}
\end{figure}
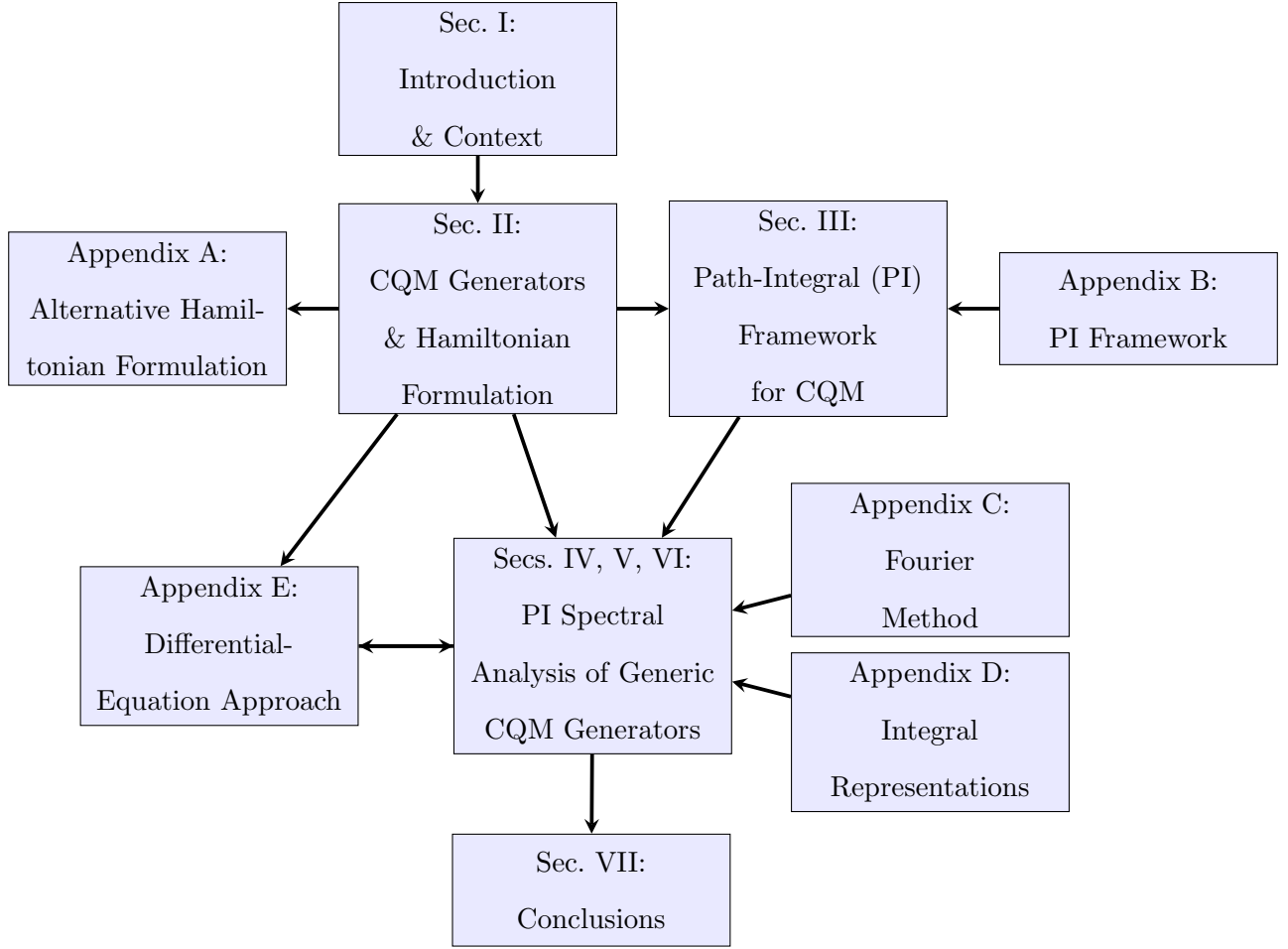

The logic of the paper organization is outlined in Fig.~\ref{fig:flowchart}. 
The flowchart stresses the need to use an appropriate, alternative Hamiltonian formulation
for the path-integral framework before the spectral analysis of the CQM generators is performed;
then, from both the Hamiltonian formulation and the path-integral treatment, all the properties 
of the CQM generators follow systematically, with additional technical details provided by the appendices.

\section{Symmetry Properties and Generators of Conformal Quantum Mechanics (CQM): Hamiltonian Formulation}
\label{sec:symmetries}

As outlined in Sec.~\ref{sec:intro}, the dAFF model is the limiting $D=1$ 
field theory of Eq.~(\ref{eq:CFT-general-Lagrangian}).
 While Eq.~(\ref{eq:CQM-ISP-Lagrangian_1d}) is the simplest form of CQM,
in this paper, we consider the more general multicomponent case, where the $d$ components 
correspond to the $d$ spatial dimensions of the position coordinates $\mathbf{Q}(t)$ in quantum mechanics,
evolving dynamically with respect to the ordinary time $t$.
This physical system, with Lagrangian
\begin{equation}
L=\frac{ M}{2} \,  \dot{\mathbf{Q}}^2 - \frac{\lambda}{2 Q^2}
\; ,
\label{eq:CQM-ISP-Lagrangian}
\end{equation}
can be described in polar hyperspherical coordinates, with the radial variable $Q= |\mathbf{Q}| \geq 0$ 
defined in the nonnegative half-line.
In the path-integral framework, a system with Lagrangian~(\ref{eq:CQM-ISP-Lagrangian}) can be more 
naturally described within a class of radial functional problems, as we will see in Sec.~\ref{sec:PI-setup_CQM}.
In Eqs.~(\ref{eq:CQM-ISP-Lagrangian_1d}) and (\ref{eq:CQM-ISP-Lagrangian}),
we are adopting a definition of the coupling constant with an additional factor of $1/2$ for convenience,
consistent with the numerical factors of Ref.~\cite{AFF:76}, and rescaled in the transition to the quantum theory,
with $\lambda = \hbar^2 \, g/M$, with $g$ dimensionless.
In addition, we are inserting an extra dimensional parameter $M$ that can be interpreted as the mass of 
a nonrelativistic particle in the conventional form of quantum mechanics. Often, the choices 
 $M=1$ and $\hbar =1$ are made for the sake of simplicity (including in Ref.~\cite{AFF:76}), but we will keep
 conventional dimensional parameters to help in some of the derivations 
 with the use of appropriate analytic continuations, limiting procedures,
 and familiar physical interpretations for both the Hamiltonian operator and the other 
 symmetry generators.

 \subsection{Conformal Symmetries}
 
Rather than starting directly from the Lagrangian~(\ref{eq:CQM-ISP-Lagrangian}),
one can begin the problem looking at the more general conditions that define  
conformal invariance for a generic Lagrangian 
\begin{equation}
L=\frac{ M}{2} \,  \dot{\mathbf{Q}}^2 - V(\mathbf{Q})\
\; .
\label{eq:CQM-general-Lagrangian}
\end{equation}
From Noether's theorem,
for systems with Lagrangians of the form~(\ref{eq:CQM-general-Lagrangian})
 the necessary conditions to yield an invariant action
 under general time reparametrizations~\cite{AFF:76,Jackiw:80,Jackiw:90}
 (with time transformations
such that the Lagrangian itself is not necessarily invariant) are:
(i) the potential function $V(\mathbf{Q})$ is a
homogeneous function of degree $ -2 $;
(ii)  the time transformations appear restricted to any of three independent building blocks:
 time translations, time dilatations, and inverse time translations.
 In what follows, $\textbf{P} \equiv \textbf{P}_{ \mathbf{Q} } $ is the momentum conjugate to $\mathbf{Q}$. 
 Then, the transformation algebra consists of the three generators (including operator ordering at the quantum level):
\begin{itemize}
\item
 The Hamiltonian
\begin{equation}
H = \frac{ \textbf{P}^2 }{2M} + V(\mathbf{Q})
\; ,
\label{eq:CQM-general-Hamiltonian}
\end{equation}
associated with time translations (with respect to $t$).

\item
The dilation generator
\begin{equation}
D= t H - \frac{1}{4} \left( \mathbf{Q} \cdot \textbf{P} + \textbf{P} \cdot \mathbf{Q}
 \right)
 \;  ,
 \label{eq:dilation-op}
\end{equation}
which enforces scale transformations.

\item
The special
conformal generator
\begin{equation}
K= H t^{2} - \frac{1}{2} (\mathbf{Q} \cdot \textbf{P} + \textbf{P} \cdot \mathbf{Q} ) \,
  t + \frac{1}{2} M Q^{2}
 \;  
\label{eq:special-conformal-op}
\end{equation}
that enforces inverse time translations.
\end{itemize}
This class of conformal Hamiltonians includes both the 
inverse square potential (ISP) $V(\mathbf{Q}) \propto Q^{-2}$ as well as contact interactions---most notably 
the two-dimensional delta~\cite{Jackiw:delta_91}, in addition to anisotropic versions of the ISP and the derivative-delta interaction in one spatial dimension~\cite{delta-prime_2021}.
While the symmetry properties of these realizations are generically the same for the whole conformal class, the analytical details of the solutions are specifically dependent on the chosen model. In this paper, we only consider the dAFF model of {\em non-contact interactions\/} that corresponds to a (0+1)-dimensional conformal field theory; in this setting, this amounts to the Lagrangian of Eq.~(\ref{eq:CQM-ISP-Lagrangian}).

The composition of the three basic types of 
transformations~(\ref{eq:CQM-general-Hamiltonian})--(\ref{eq:special-conformal-op})
generates the linear fractional transformation 
\begin{equation}
\tilde{t} =\frac{ \alpha t +\beta}{\gamma t + \delta} 
\; ,
\label{eq:symm-time-transf}
\end{equation}
governed by a matrix in the special linear group $SL(2,\mathbb{R})$
(with $\alpha \delta - \beta \gamma = 1$),
and with transformed field variable 
\begin{equation}
\tilde{ \mathbf{Q} }(\tilde{t}) = \frac{ \mathbf{Q} (t) }{ \gamma t + \delta }
\; .
\label{eq:symm-Q-transf}
\end{equation}
At the quantum level, Eq.~(\ref{eq:symm-Q-transf}) is written in the Heisenberg picture, and has the same form
as its classical counterpart.
The associated symmetry group is $SL(2,\mathbb{R})$, which is homomorphic to  $SO(2,1) $, with the generators
 $\{H, D, K\}$ satisfying the commutator relations~\cite{AFF:76}
 \begin{equation}
 [D,H]= -i \hbar H
\; , \qquad 
[D,K]= i \hbar K
\; , \qquad 
  [H,K]=2 i \hbar D
  \; ,
\end{equation}
which correspond to the one-dimensional case 
of the algebra of the generic conformal group.
To introduce an appropriate Cartan-Weyl basis~\cite{wyb:74}, 
one can replace $ H$ and $K$ by the operators $R$ and $S$ defined by the linear combinations
\begin{align}
 R = & \frac{1}{2} \left( \frac{1}{a} K + a H \right)
\label{eq:operator-R}
\; ,
\\
 S \equiv - S' = & \frac{1}{2} \left( \frac{1}{a} K - a H \right)
\; ,
\label{eq:operator-S}
\end{align}
where $a$ is an arbitrary parameter with dimensions of time
(with $R$ and $S$, as well as $D$, defined with dimensions of action).
The commutators of the Cartan-basis generators are 
$[S,D] = - i \hbar R$,
$[D,R] = i \hbar S$, and
$[R,S] = i \hbar D$.
The $so(2,1) $ algebra has rank one and dimension 3, 
with the Cartan generator $R$ and the ladder operators $L_{\pm} $
defined below, along with their corresponding commutator relations:
\begin{align}
& L_{\pm} = S \pm i D 
\\
& [\hbar^{-1} R, L_{\pm}]= \pm L_{\pm}
\; , \qquad 
[L_{+}, L_{-}]= - 2 \hbar R
\; .
\end{align}
In this paper, we will use the operator $S'=-S$ (with the sign reversed, as defined above), 
so that a more transparent physical interpretation can be given to the corresponding 
effective potential, as in Sec.~\ref{sec:spectral_CQM_hyperbolic}---this choice 
can be found in the literature, for example, in Ref.~\cite{Arzano-1}.

\subsection{Generalized Conformal Generators: Dynamics and Hamiltonian Formulation} 
\label{subsec:conformal-generator-G} 

The properties of the conformal generators are well-known from the exhaustive analysis of Ref.~\cite{AFF:76}.
The operator $R$ generates a compact subgroup [two-dimensional rotations as the $O(2)$ subgroup of $SO(2,1)$], 
and $S$ and $D$ generate non-compact boosts.
These distinct behaviors can be characterized
by considering the generalized generator~\cite{AFF:76}
\begin{equation}
G= uH + vD + w K
\; ,
\label{eq:gen-generator}
\end{equation}
with discriminant 
\begin{equation}
\Delta = v^2 - 4uw
\; 
\label{eq:gen-generator_discriminant}
\end{equation}
that determines its nature:
rotations (elliptic type), for $\Delta <0$, which include $R$;
boosts (hyperbolic type), for $\Delta >0$, 
which include both $S$ and $D$;
 and parabolic (``lightlike'') operators, for $\Delta =0$,
 which include the original $H$ and $K$ operators.

A complete analysis~\cite{AFF:76}
 in terms of the operators~(\ref{eq:gen-generator}) shows that $G$
can be regarded as the generator of time evolution with respect to a modified, effective time $\tau$.
For the $d$-component field ${\mathbf{Q} } (t)$,
 from the definitions~(\ref{eq:CQM-general-Hamiltonian})--(\ref{eq:special-conformal-op})
and (\ref{eq:gen-generator}),
the basic equation satisfied in the Heisenberg picture by the action of the operator $G$
on the field variables ${\mathbf{Q} } (t) $ is
\begin{equation}
 \frac{1}{i \hbar}  \left[ {\mathbf{Q} } (t) , G \right] = 
f_{G}(t)  \frac{ d {\mathbf{Q} } (t) }{ d t }
-
\frac{1}{2} \frac{ d f_{G}(t)}{dt} \,   {\mathbf{Q} } (t) 
\; ,
\label{eq:action-G-on-Q-t}
 \end{equation}
 where 
\begin{equation}
f_{G} (t) = u + vt + wt^{2} = \sigma \, \left| u + vt + wt^{2} \right|
\; ,
\label{eq:f(t)}
\end{equation}
with the sign $ \sigma \equiv \sigma_{G} = \mathrm{sgn} [f_{G}(t)] $ allowing for
arbitrary linear combinations of the generalized generator~(\ref{eq:gen-generator}).
The dynamical equation~(\ref{eq:action-G-on-Q-t}),
just as in the $d=1$ case~\cite{AFF:76},
can be simplified to a more standard form by redefining the dynamical time to $\tau$ according to
\begin{equation}
 d \tau =  \frac{ dt }{ u + vt + wt^{2}  }
\label{eq:G_tau}
\; ,
\end{equation}
and the field variables to 
\begin{equation}
 {\mathbf{r} } (\tau) 
 = \frac{ {\mathbf{Q} } (t) }{ \left| u + vt + wt^{2} \right|^{1/2} }
 \; .
 \label{eq:G_Q-field}
\end{equation}
In Eq.~(\ref{eq:G_Q-field}), the absolute value is needed to reproduce 
Eq.~(\ref{eq:action-G-on-Q-t}), with the auxiliary Eq.~(\ref{eq:f(t)}).
Equation~(\ref{eq:G_tau}) can be easily integrated; particular cases (including $R$, $S$, $D$, and $K$)
are listed in Ref.~\cite{AFF:76}.
Thus, the dynamics in the Heisenberg picture is fully described by the equation
\begin{equation}
 \frac{1}{i \hbar}  \left[ {\mathbf{r}  }(\tau ) , G \right] = \frac{ d {\mathbf{r} } (\tau) }{ d \tau }
 \; ,
 \label{eq:action-G-on-q-tau}
\end{equation}
which shows that $G$ acts as an effective Hamiltonian for time evolution.
One can alternatively describe the dynamics in a generalized Schr\"{o}dinger picture
with respect to $G$, 
in terms of evolving state vectors
$ \ket{ \Psi (\tau) }_{_{\!  \!(S)}}  \equiv \ket{ \Psi (\tau) }
= U_{G} (T) \,  \ket{ \Psi (\tau_{0}) }$, where 
$\ket{ \Psi (\tau_{0}) }$ is the original state in the Heisenberg picture
and 
\begin{equation}
 U_{G}(T)  \equiv U_{G}(\tau; \tau_{0}) 
 = \exp \left[ -\frac{i}{\hbar} G  \, ( \tau - \tau_{0} ) \right]
 \; 
 \label{eq:tau-evolution-op}
\end{equation}
acts as an effective evolution operator; thus, the state 
vector $ \ket{ \Psi (\tau)  }$ satisfies the generalized  Schr\"{o}dinger equation
\begin{equation}
  G \ket{ \Psi (\tau)  }
   = i  \hbar \frac{ d \ket{ \Psi (\tau)  } }{ d \tau }
 \; .
 \label{eq:action-G-on-Psi-tau}
\end{equation}
  Then,
  Eqs.~(\ref{eq:action-G-on-q-tau})
and (\ref{eq:action-G-on-Psi-tau})
describe the evolution of the field variables and states under the action of $G$ in the Heisenberg and
 Schr\"{o}dinger pictures respectively,
with the new effective time $\tau$; in other words, $G$ formally acts as $i \hbar \partial_{\tau}$.
  In these transformed variables, the time evolution only depends on the effective time difference 
  $T = \tau - \tau_{0} $.
It should be noted that this is {\it not\/} the original Schr\"{o}dinger picture associated with the Hamiltonian $H$,
and it is $G$-specific.

In principle, the analysis leading to Eqs.~(\ref{eq:G_tau})--(\ref{eq:action-G-on-Psi-tau})
 allows a reformulation of the theory in terms of a new ``Hamiltonian'' 
$H_{G} (\mathbf{r}, \mathbf{p} ) \equiv G$.
This assignment $H_{G} \equiv G$, by abuse of notation, simply states that the value of $G$ 
is to be rewritten in terms of the transformed canonical variables
$(\mathbf{r}, \mathbf{p} )$.  
By construction, ${\mathbf{r} } $ is given by Eq.~(\ref{eq:G_Q-field}), but $\mathbf{p} $ has to be 
defined consistently in such a way that
$| f_{G} |^{1/2} d { \mathbf{r} }/d \tau  = f_{G} \, d { \mathbf{Q} }/dt   - \dot{ f_{G} }  \, \mathbf{Q}/2 $
(with the dot notation for derivatives with respect to the original time $t$)---an expression that
can be read off by comparison of
Eqs.~(\ref{eq:action-G-on-Q-t}) and (\ref{eq:action-G-on-q-tau}) or from Eqs.~(\ref{eq:G_tau})--(\ref{eq:G_Q-field}).
An obvious choice for $ \mathbf{p} $ is the one that satisfies
$ \mathbf{p} = M \dot{ \mathbf{r} }$, whence the transformed momentum is given by
\begin{equation}
\mathbf{p} =  \sigma \, | f_{G} |^{1/2} 
 \biggl(
 \mathbf{P} - \frac{\dot{ f_{G}} }{ 2 f_{G} } \,  M \mathbf{Q} 
 \biggr)
 \; .
 \label{eq:momentum-p_q}
 \end{equation}
  From
Eq.~(\ref{eq:gen-generator}), this implies that the operator
$G= uH + v(tH + D_{0}) + w(t^{2} H+ 2 t D_{0} +K_{0} ) = f_{G} H + \dot{ f_{G} } D_{0} + w K_{0}$
[where $ D_{0}$ and $ K_{0}$ are the time independent parts in 
Eqs.~(\ref{eq:dilation-op}) and (\ref{eq:special-conformal-op})] takes the ``normal'' Hamiltonian form
\begin{equation}
G \equiv H_{G} (\mathbf{r}, \mathbf{p} )
=
\sigma \, \tilde{H}_{G} (\mathbf{r}, \mathbf{p} )
\; ,
\label{eq:G-as-Hamiltonian} 
\end{equation}
where
\begin{equation}
\tilde{H}_{G} (\mathbf{r}, \mathbf{p} )
=
\frac{1}{2M} 
 p^{2}  + \frac{1}{2} \frac{\lambda}{r^{2}} + \frac{M}{2} \left(-\frac{\Delta}{4} \right) r^{2}
\; .
\label{eq:G-as-Hamiltonian_tilde} 
\end{equation}
In Eq.~(\ref{eq:G-as-Hamiltonian_tilde}), the usual convention $A= |\mathbf{A}|$ for spatial vectors is used,
and the discriminant~(\ref{eq:gen-generator_discriminant}) is shown to be the same as $\Delta = \dot{ f_{G} }^{2} - 4w f_{G}$.
The generalized momentum $ \textbf{p} $ in Eq.~(\ref{eq:momentum-p_q}) and the
 functional form of $H_{G} (\mathbf{r}, \mathbf{p} ) $ 
 in Eqs.~(\ref{eq:G-as-Hamiltonian}) and (\ref{eq:G-as-Hamiltonian_tilde}) can also
 be verified operationally at the classical level by a straightforward computation of the Hamiltonian from the transformed 
 Lagrangian $L_{G}$, using Eqs.~(\ref{eq:CQM-ISP-Lagrangian}) and (\ref{eq:G_tau})--(\ref{eq:G_Q-field}),
 as shown in Ref.~\cite{AFF:76}.

\subsection{Equivalence Classes of the Generalized Generators and Hamiltonian $H_{G}$}
\label{subsec:equivalence-classes}

Equations~(\ref{eq:G-as-Hamiltonian})--(\ref{eq:G-as-Hamiltonian_tilde}) establish the Hamiltonian formulation
of the generalized generators $G$, which are initially defined on the Lie algebra $so(2,1)$. 
The Hamiltonian $H_{G}$ is generated by a surjective mapping that endows them with a 
functional form with respect to the transformed canonical variables~(\ref{eq:G_Q-field}) and (\ref{eq:momentum-p_q}),
such that its values only depend on the discriminant $\Delta$.
As a result, 
there is a continuous range of linear combinations $G$, 
with values of $u$, $v$, and $w$ with the same discriminant, that give rise to the same effective Hamiltonian 
$H_{G} (\mathbf{r}, \mathbf{p} )$.
Moreover, the generators $G$ define equivalence classes 
via the relation $G_{1} \simeq G_{2}$ iff $H_{G_{1}} =H_{G_{2}}$,
which identifies them by the same values of $\Delta$ 
(i.e., $\Delta_{1} = \Delta_{2}$). 
This equivalence relation produces surprising connections between 
apparently distinct operators, such as, for example, $H$ and $K$, or $D$ and $S'$.
The corresponding effective Hamiltonians are identical and
yield equivalent models within the theory, with identical dynamical evolutions with respect to $\tau$, 
as well as equivalent spectral properties.
Therefore, for our main purposes,
 we can assume that $G$ and $H_G$ are identified without any further qualifications:
the Hamiltonian/Lagrangian frameworks directly lead 
to the path integral and the spectral properties, and the practical use of an
equal sign (instead of $\simeq$) is fully justified for the remainder of this paper.
 
 In summary,
 the theory is organized in equivalence classes labeled by the value of the discriminant $\Delta$. 
Moreover, these classes are further categorized in the three types of operator behavior:  
$\Delta \lessgtr 0$ and $\Delta =0$ (elliptic, hyperbolic, and parabolic),
 whose distinct properties can be understood from the change in sign of the
quadratic potential in Eq.~(\ref{eq:G-as-Hamiltonian_tilde}).
These behaviors define a coarser equivalence relation leading to the three distinct 
broad classes of operators.

Regarding the spectral characterization of operators that we are addressing in this paper,
the primary goal is to fully determine the eigenstates and eigenvalues of the generalized conformal generator $G$.
We denote its eigenvalues as $\hbar \mathfrak{g}$, where
$\mathfrak{g} $ is dimensionless; then, from Eq.~(\ref{eq:G-as-Hamiltonian}), 
it follows that 
\begin{equation}
\hbar \mathfrak{g} = {E}_{G}  = \sigma \tilde{E}_{G} 
\label{eq:eigen-g_E}
\; ,
\end{equation}
in terms of the eigenvalues of the associated $\tilde{H}_{G} (\mathbf{r}, \mathbf{p} )$.
(It should be noted that, in Ref.~\cite{AFF:76}, the eigenvalues of $G$ are denoted by $G'$.)
In light of Eq.~(\ref{eq:eigen-g_E}), the eigenstates will be labeled interchangeably with 
$ \mathfrak{g}$ or ${E}_{G}$, and the latter may be simplified as $E$ if there is no obvious conflict.

Equation~(\ref{eq:G-as-Hamiltonian_tilde}) 
is an exact analog of a standard quantum-mechanical central problem with Hamiltonian and potential 
\begin{equation}
\tilde{H} (\mathbf{r}, \mathbf{p} )
 =  \frac{p^2}{2M} + \tilde{V}(r)
 \; , \; \; \; \; \; \;
 \tilde{V} (r) = \frac{1}{2} M \omega^2 r^2 + \frac{\hbar^2}{2M} \, \frac{g}{r^{2}}
\label{eq:potential_HO+ISP}
 \; ,
\end{equation}
 which can be interpreted as representing a harmonic oscillator of squared ``frequency''
\begin{equation}
\omega^{2} = -\frac{\Delta}{4 }
\; 
\label{eq:squared-frequency} 
\end{equation}
(including possibly imaginary frequencies)
superimposed with the original inverse square potential.
Even though our conformal problem has a different meaning, this analog parametrization is insightful and the
``CQM frequency'' from Eq.~(\ref{eq:squared-frequency}) will play an important role in describing the 
spectral solutions.
In what follows, we will also use the functional notation
  $\tilde{H} \bigl[ M, g, \omega \bigr] \equiv \tilde{H}_{\mathbf{r}, \mathbf{p} } \bigl[ M, g, \omega \bigr] $
   for the corresponding parameter dependence, where 
   $ \tilde{H}_{\mathbf{r}, \mathbf{p} }  \equiv  \tilde{H} (\mathbf{r}, \mathbf{p} )$, with the appropriate canonical variables.
   A summary of the different Hamiltonian notations used in the main text is given in Table~\ref{tab:h_notations}.
\begin{table}[!htbp]
    \centering\setcellgapes{4pt}\makegapedcells \renewcommand\theadfont{\normalsize\bfseries}
    \caption{Here we tabulate the different notations that we have used for specific Hamiltonians. All the expressions are written with position and momentum arguments denoted by $\mathbf{r}$ and $\mathbf{p}$, respectively.\\~}
    \begin{tabular}{|c|c|}
        \hline Notation & Expression \\
        \hline\hline $H(\mathbf{r,p})$ & $\frac{p^2}{2M} + \frac{g}{r^2}$ \\
        \hline $\tilde{H}_G(\mathbf{r,p})$ & $\frac{p^2}{2M} + \frac{\lambda}{2r^2} + \frac{M}{2}\left(-\frac{\Delta}{4}\right) r^2$\\
        \hline $G \equiv H_G$ & $\sigma\tilde{H}_G(\mathbf{r,p})$ \\
        \hline $\tilde{H}(\mathbf{r,p}) \equiv \tilde{H}_{\mathbf{r,p}} \equiv \tilde{H}[M,g,\omega]$ & $\frac{p^2}{2M} + \frac{1}{2}M\omega^2 r^2 + \frac{\hbar^2}{2M} \frac{g}{r^2}$\\
        \hline $\hat{H}$ & Generic quantum mechanical Hamiltonian\\
        \hline
    \end{tabular}
    \label{tab:h_notations}
\end{table}

The Hamiltonian function of Eq.~(\ref{eq:G-as-Hamiltonian_tilde}) exhibits a scaling property 
that is a consequence of the definition (\ref{eq:gen-generator}), which is linear in the coefficients $u, v, w$, 
and with $\Delta$ given as a quadratic function~(\ref{eq:gen-generator_discriminant}).
If this Hamiltonian function is rewritten in the generic analog form of Eq.~(\ref{eq:potential_HO+ISP}),
with frequency~(\ref{eq:squared-frequency}), then it satisfies the scaling
\begin{equation}
\tilde{H} \bigl[ M, g, c \,  \omega \bigr]
=
 c 
\tilde{H} \bigl[ M, g,  \omega \bigr]
\; ,
\label{eq:tilde-H_scaling-property} 
\end{equation}
where $c>0$ is an arbitrary real and positive scaling constant.
In other words, the Hamiltonian
 is a homogenous function of first degree with respect to the frequency $\omega$.
An outline of the proof is as follows: if $G$ is rescaled with the coefficients $u'= c u$, $v'= c v$, $w'= c w$, then it takes
the value $G'= cG=  \sigma c 
\tilde{H} \bigl[ M, g,  \omega \bigr]$, without a change in $\sigma$, but
with the discriminant changing as $\Delta' = c^2 \Delta$.
At the same time, Eq.~(\ref{eq:squared-frequency}) gives
 $\omega' = c \, \omega$ and this yields 
 $G' = \sigma \tilde{H} \bigl[ M, g, c \,  \omega \bigr]$,
 whence
 Eq.~(\ref{eq:tilde-H_scaling-property}) follows by comparing the two expressions for $G'$.
This can be also be shown more generally directly at the level of Eq.~(\ref{eq:potential_HO+ISP}) due to the general scaling properties of this Hamiltonian (as in ordinary dimensional analysis), in terms of the frequency parameter, with $\tilde{H}$ and $\omega$ having the same scaling.

From Eq.~(\ref{eq:G-as-Hamiltonian_tilde}),
the spectral properties of the different types of conformal generators can be immediately understood.
In this procedure, we can choose the prototypical generators $R$, $H$, and $S'$ as representatives of the three 
main classes of operator behaviors: their general properties are shared by the generators $G$ with 
$\Delta <0$, $\Delta =0$, and  $\Delta >0$, respectively.
Then,
 $\omega  = \sqrt{ | \Delta | }/2$
  for $\Delta < 0$
and $\omega  = -i \sqrt{ \Delta }/2$   for $\Delta > 0$ 
(the choice of sign for the latter is discussed in Sec.~\ref{sec:spectral_CQM_hyperbolic}),
with $\omega = 0 $ for $\Delta = 0$. 
Moreover, 
we can spell out the parameter dependence of all the generators, starting with the prototypical operators
$R$, $H$, and $S'$, whose equivalence classes have discriminants
 $\Delta = -1$,  $\Delta = 0$, and $\Delta = 1$, respectively. 
 Thus,
\begin{align}
& R = 
\tilde{H} \bigl[ M, g, \omega=1/2 \bigr]
 \; ,
\label{eq:operator-R_parameters}
\\
& H = \tilde{H}  \bigl[ M, g, \omega= 0 \bigr] 
 \; ,
\label{eq:operator-H_parameters}
\\
& S' = \tilde{H}  \bigl[ M, g, \omega=-i/2 \bigr] 
 \; .
\label{eq:operator-S_parameters}
\end{align}
In addition, as in Ref.~\cite{AFF:76}, the parameters $\hbar$ and $M$ could be set equal to unity 
(though we will mostly keep the general parametrization for convenience).
Most importantly, applying the scaling~(\ref{eq:tilde-H_scaling-property}), it follows that
$ \tilde{H}_{G} = \sqrt{| \Delta |} \, \tilde{H} \bigl[ M, g, \omega= 1/2 \bigr] $ 
for elliptic operators, and
$ \tilde{H}_{G} = \sqrt{ \Delta } \, \tilde{H} \bigl[ M, g, \omega= -i/2 \bigr] $ 
for hyperbolic operators.
These scaling relations 
lead to the following complete characterization of the
 three classes of operators.
\begin{itemize}
\item
Elliptic Generators, defined by $\Delta<0$. 
These are ``generalized $R$ operators,'' of the form
\begin{equation}
  G  =  \sigma \sqrt{| \Delta |}  \; R
\; ,
\label{eq:elliptic-generators} 
\end{equation}
where $R$ is the prototypical elliptic operator
($\Delta =-1$)  
of Eqs.~(\ref{eq:operator-R}) and (\ref{eq:operator-R_parameters}).

\item
Parabolic Generators, defined by $\Delta=0$. 
These are ``generalized $H$ operators,'' of the form
\begin{equation}
 G = \sigma  H
\; ,
\label{eq:parabolic-generators} 
\end{equation}
where $H$ is the prototypical parabolic operator 
of Eqs.~(\ref{eq:CQM-general-Hamiltonian}) and (\ref{eq:operator-H_parameters}).

\item
Hyperbolic Generators, defined by $\Delta>0$. 
These are ``generalized $S$ operators,'' of the form
\begin{equation}
G = \sigma \sqrt{ \Delta }  \; S'
\; ,
\label{eq:hyperbolic-generators} 
\end{equation}
where $S'$ is the prototypical hyperbolic operator 
($\Delta =1$)  
of Eqs.~(\ref{eq:operator-S}) and (\ref{eq:operator-S_parameters}).
 
\end{itemize}
This analysis verifies that the theory is organized in equivalence classes defined by the value of $\Delta$, and where 
$\sigma = \text{sgn} ( f_{G} )$ just gives the ``orientation'' of the operator spectrum. Within each class,
regardless of the values of $u$, $v$, and $w$, the effective Hamiltonian, dynamical evolution, and spectral 
properties are identical (except for a possible rescaling of the eigenvalues)---but the 
form of the effective time $\tau$ (in terms of $t$) is specific to each operator.
It is also noteworthy that 
the results of Eqs.~(\ref{eq:elliptic-generators})--(\ref{eq:hyperbolic-generators})
are completely general, and apply even for the parabolic operator $K$ (which, according to this, gives
the same propagator as $H$) and for the hyperbolic operator $D$
(having the same propagator as $S'$). 
Incidentally, an alternative form of this theory is 
presented in Appendix~\ref{app:generators-Hamiltonian_dimensional-rep}, which is restricted to the cases with
$u \neq 0$ only, but has some advantages in terms of dimensional analysis.

In this section,
we have established the general framework for the generators $G$.
Next, we will use the parametrization of 
Eqs.~(\ref{eq:G-as-Hamiltonian})--(\ref{eq:G-as-Hamiltonian_tilde})
for a complete path-integral analysis of the spectral properties of all the
 conformal operators, for three associated operator classes (elliptic, parabolic and hyperbolic).

\section{Path-Integral Framework: Basic Setup for Conformal Generators} 
\label{sec:PI-setup_CQM}

All properties of a physical system can be completely characterized by using path-integral methods. 
This includes the spectral properties of operators relevant to the system. We will show how to derive such
properties for the CQM operators $R$, $H$, and $S$, and, by extension, 
according to Eqs.~(\ref{eq:elliptic-generators})--(\ref{eq:hyperbolic-generators}),
for the generalized generators $G$ of Eq.~(\ref{eq:gen-generator}).
Due to the form of the Lagrangian~(\ref{eq:CQM-ISP-Lagrangian}) of the dAFF model 
and the special conformal operator~(\ref{eq:special-conformal-op}), 
leading to the effective Hamiltonian~(\ref{eq:G-as-Hamiltonian_tilde}),
 we need the general $d$-dimensional path-integral solution for the radial harmonic oscillator,
 along with the inverse square potential, i.e., for a central problem with Hamiltonian and potential given by Eq.~(\ref{eq:potential_HO+ISP}).
Of course, this is strictly needed for the multicomponent case of CQM, but it is also a useful
approach to properly incorporate the inverse square potential with an appropriate analytic continuation 
(see below and Appendix~\ref{app:PI-setup}), even in the simplest case $d=1$.

Remarkably, we are computing the path-integral propagators associated with the operators
$R$, $H$, and $S$, as well as their generalized forms $G$, as {\em generators of dynamical evolution\/}. 
The specific functional form of these generators involves the generalized canonical variables defined in 
Eqs.~(\ref{eq:G_Q-field}) and (\ref{eq:momentum-p_q}) (or rescalings thereof).
For $H$ itself, the effective propagator is indeed the generator of dynamics in ordinary time $t$, but all the other 
generators involve dynamical evolutions with respect to their ``natural'' effective times $\tau$. 
(In Appendix~\ref{app:generators-Hamiltonian_dimensional-rep},
an alternative, dimensional framework is defined for the cases with $u \neq 0$, in terms
 of a rescaled time $\tilde{\tau}$, which plays a similar role.)
Moreover, in addition to describing the dynamics, the propagators play the role of
{\it generating functions for the spectral decomposition\/} and will allow us to find the spectral data, 
namely, the eigenvalues and eigenstates for the family of conformal generators.

\subsection{Path-Integral Framework}
\label{subsec:PI_framework}

We begin setting up the framework with some notational remarks.
In the theory outline of this subsection and in Appendix~\ref{app:PI-setup}
we typically denote by $t$ the dynamical time, with 
$\hat{H}$ describing a generic Hamiltonian, as in the usual quantum-mechanical applications.
Within this general framework, we use a radial harmonic oscillator (properly generalized to include inverted oscillators 
and any number of dimensions) as an analog system for a generic conformal generator $G$. 
In this sense, when describing the spectral properties and dynamics generated by $G$, 
the general framework involves replacing $\hat{H}$ by $G$ or $\tilde{H}$, 
and $t$ by $\tau$.

Thus, our main focus will be on
a generic Hamiltonian 
$\hat{H}=p^2/2M + V({\bf r})$, with a 
 time-independent and central potential
$V({\bf r})= V(r)$ in  $d$ spatial dimensions.
We begin our construction by considering the quantum-mechanical propagator 
\begin{equation} 
 K_{(d)}  ( {\bf r}'', {\bf r}' ; t'',t' ) 
=
\left\langle 
{\bf r}''
\left| 
\hat{T}
\exp \left[
 -\frac{i}{\hbar} \int_{t'}^{t''}
\hat{H} dt 
\right]
\right|
{\bf r}'
 \right\rangle   
\; ,
\label{eq:propagator_QM_from_evolution_operator}
\end{equation}
for a particle of mass $M$ subject to the Hamiltonian $\hat{H}$.
 In Eq.~(\ref{eq:propagator_QM_from_evolution_operator}),
 $\hat{T}$ is the time-ordering operator.
The corresponding path-integral expression
\begin{equation}
K_{(d)}({\bf r}'', {\bf r}' ; t'',t') 
= 
\int_{  {\bf r} (t')  = {\bf r}'  }^{  {\bf r} (t'')  = {\bf r}'' }
 \;  
{\cal D} {\bf r} (t) \,
\exp \left\{ 
\frac{i}{\hbar} 
S \left[ {\bf r}(t)  \right]  ({\bf r}'', {\bf r}' ; t'',t')  
\right\}
\;  
\label{eq:propagator_QM}
\end{equation}
involves the classical action functional
$S \left[ {\bf r}(t)  \right]  ({\bf r}'', {\bf r}' ; t'',t')  $ 
for ``paths''
$ {\bf r}(t)  $ connecting the end points
${\bf r} (t')  = {\bf r}',  {\bf r} (t'')  = {\bf r}'' $.
For time-independent potentials,
the propagator $K_{(d)}({\bf r}'', {\bf r}' ; t'',t') \equiv K_{(d)}({\bf r}'', {\bf r}' ; T) $
is a function of $T=t''-t'$ alone and not of the individual endpoint times; 
we will use this assumption for the remainder of the paper.
Specifically, for the generic conformal generator $G$, the associated propagator is given by
$K_{(d)}^{(G)}  ({\bf r}'', {\bf r}' ; T)  = 
\left\langle 
{\bf r}''
\left| 
U_{G}(T)
\right|
{\bf r}'
 \right\rangle 
 $, with $U_{G}(T)$ as in Eq.~(\ref{eq:tau-evolution-op}),
 in terms of the effective time $\tau$.

In Appendix~\ref{app:PI-setup}, 
we review the required construction for central potentials in hyperspherical coordinates, 
and specifically for the radial harmonic oscillator. 
In general, the propagator $ K_{(d)}  ( {\bf r}'', {\bf r}' ; T ) $
can be formally rewritten in hyperspherical polar coordinates~\cite{Cam_DT1,Cam_DT2,Erdelyi_HTF2},
which yield a complete set of angular functions, the
$d$-dimensional hyperspherical harmonics
$Y_{l \boldsymbol{m}} ({\bf \Omega})$, 
where the quantum labels correspond to the $d$-dimensional angular momentum 
with numbers
$l$ and $\boldsymbol{m}$, and such that the set $\boldsymbol{m}$ takes 
 $g_{l}= (2l+d-2)(l+d-3)!/l!(d-2)!$
 values as multiplicity for a given $l$ (Chap.~XI in Ref.~\cite{Erdelyi_HTF2}).
The corresponding partial wave expansions of $ K_{(d)}  ( {\bf r}'', {\bf r}' ; T ) $ read
\begin{align}
& K_{(d)}({\bf r}'', {\bf r}' ; T) 
 = 
\left( r'' r' \right)^{-(d-1)/2}
\,
\sum_{l= 0}^{\infty}
\sum_{m=1}^{g_{l}}
Y_{l \boldsymbol{m}} ({\bf \Omega}'')
Y_{l \boldsymbol{m}}^{ *} ({\bf \Omega}')
\,
K_{l +\nu}(r'',r';T)
\label{eq:propagator_partial_wave_exp_Y}
\\
& =
\frac{\Gamma (\nu)}{2 \pi^{d/2} }
\,
\left( r'' r' \right)^{-(d-1)/2}
\,
\sum_{l= 0}^{\infty}
\,
(l+\nu)
\,
C_{l}^{(\nu)} (\cos \psi_{ {\bf \Omega}'', {\bf \Omega}' } ) 
\,
K_{l +\nu}(r'',r';T)
\;  ,
\label{eq:propagator_partial_wave_exp}
\end{align}
where $\nu =d/2 -1$,
 $\cos \psi_{ \, {\bf \Omega}'', {\bf \Omega}' } = {\bf \hat{r}}'' \cdot {\bf \hat{r}}' $
  (with $\hat{\mathbf{r} } = \mathbf{r}/r$), 
  and the addition theorem for hyperspherical harmonics provides the expression in terms of
Gegenbauer polynomials $C_{l}^{(\nu)}(x)$.
In Eqs.~(\ref{eq:propagator_partial_wave_exp_Y}) and (\ref{eq:propagator_partial_wave_exp}), and subsequent 
expressions, we will use the radial-coordinate symbol $r = | \mathbf{r} | $.
For central potentials,
the radial propagator $K_{l+\nu}(r'',r';T)$ is independent of the angular coordinates and quantum numbers 
$\boldsymbol{m}$.
The chosen normalization of radial prefactors is such that 
the radial propagator $K_{l+\nu}(r'',r';T)$ satisfies the composition property
\begin{equation}
K_{l+\nu}(r'',r';t''-t')
=
\int_{0}^{\infty}
dr
\,
K_{l+\nu}(r'',r;t''-t)
K_{l+\nu}(r,r';t-t')
\;  . 
\end{equation}
For the radial propagator associated with $G$ as generalized Hamiltonian with effective time $\tau$, the notation 
$K^{(G)}_{l+\nu}(r'',r';T)$ will be used for the remainder of the paper.

An explicit expression for the radial propagator 
in Eqs.~(\ref{eq:propagator_partial_wave_exp_Y})--(\ref{eq:propagator_partial_wave_exp}) 
can be derived in the time-sliced path integral.
In short, for a time lattice $t_{j}= t'+j \epsilon$,
for the time interval $T=t''-t'$ corresponding to the
 end points
${\bf r}_{0} \equiv {\bf r'}$
and 
${\bf r}_{N} \equiv {\bf r''}$;
in this lattice,
$\epsilon=
T/N$, with $j=0, \cdots,N$,
with $t_{0} \equiv t'$ and $t_{N} \equiv t''$,
such that $ {\bf r}_{j} = {\bf r} (t_{j})$.
Then, with the restriction to the half-line $r(t) \geq 0$,
 the radial propagator admits a formal continuum-limit 
 representation~\cite{Grosche-PI, Grosche-Steiner_PI-QM,Kleinert-PI, Grosche:1987,Inomata_PIs-1992}
\begin{equation}
K_{l+\nu}(r'',r';T)
=
\int
{\mathcal D}  r(t)  \,
\; 
w_{l+ \nu} [r^{2}]
\exp 
\left\{
\frac{i}{\hbar} 
\int_{t'}^{t''}
dt
\left[
\frac{M}{2} \,
\dot{r}^{2}
-
V(r)
\right]
\right\}
\;   ,
\label{eq:propagator_QM_spherical_coords_expansion_continuum}
\end{equation}
in terms of the usual one-dimensional path-integral measure 
${\mathcal D}  r(t)  $ (as in Cartesian coordinates), but with a nontrivial radial functional weight 
$w_{l+\nu} [r^{2}] = \lim_{N \rightarrow \infty} 
w_{l+\nu}^{(N)} [r^{2}]
$
[see Eq.~(\ref{eq:Besselian-functional-weight})],
which involves Bessel functions of order $l+\nu$.
This has been called a Besselian path integral~\cite{Grosche-PI, Grosche-Steiner_PI-QM}.

The general expressions for Besselian path integrals are given in Appendix~\ref{app:PI-summary}, 
including statements on the associated property  of
interdimensional dependence~\cite{interdimensional, Fischer:1992}.
This framework leads to the theorem for the insertion of inverse square potential terms as generalized angular momenta, as proved in Refs.~\cite{Peak-Inomata:1969, Inomata_PIs-1992, Fischer:1992}.
 More precisely:
 \begin{theorem*} 
Given a potential 
 $\tilde{V} (r) = V (r) + (\hbar^2/2M)gr^{-2}$,
 the propagator
 $\tilde{K}_{l+\nu} \equiv K_{l+\nu} [ \tilde{V} ] $ 
 associated with $\tilde{V} (r)$ 
 in Eq.~(\ref{eq:propagator_QM_spherical_coords_expansion_continuum})
 is equivalent to the propagator $K_{\mu} [ V ] $ with the reduced potential $V(r)$ and
 an effective, typically non-integer, angular momentum variable
\begin{equation}
 \mu
 = \sqrt{(l+ \nu)^2 + g}
 \label{eq:ang-momentum_ISP-extension}
 \; ,
 \end{equation}
such that 
\begin{equation}
\tilde{K}_{l+\nu} \equiv K_{l+\nu}  [ \tilde{V} ] 
= K_{\mu} [ V ] 
\; .
\label{eq:propagator_with-ISP}
\end{equation}
   \end{theorem*}
In short, this procedure
  expresses a practical rule for the insertion of inverse square potential terms 
  by the straightforward analytic continuation~(\ref{eq:ang-momentum_ISP-extension}),
  which amounts to the introduction of a continuous, effective angular momentum
 $  l_{\mathrm{eff}} = \mu -  \nu  $.
 As a result of this asymptotic recombination (see Appendix~\ref{app:PI-summary}), 
 the theorem defines the potential $\tilde{V}$ to be
 used in Eq.~(\ref{eq:potential_HO+ISP}).

\subsection{Propagator for Generalized Radial Harmonic Oscillator}
\label{subsec:propagator_radial-HO}

The derivation of the general path integral 
for the radial propagator of the $d$-dimensional harmonic oscillator involves
explicit use of the radial path integral~(\ref{eq:propagator_QM_spherical_coords_expansion_continuum}).
 With the inclusion of an additional inverse square potential,
i.e., for a Hamiltonian $\tilde{H} (\mathbf{r}, \mathbf{p} )$ with
an effective potential of the form~(\ref{eq:potential_HO+ISP}), the propagator takes the form
\begin{equation}
K^{(RHO)}_{l+\nu} (r'',r';T)  
=
\frac{M \omega }{i \hbar \sin \omega T} 
\,
\sqrt{ r' r''} 
\,
\exp \left[ \frac{i M \omega }{2 \hbar } \left(r'^{2} + r''^{2} \right) \cot \omega T \right]
I_{\mu} \left( \frac{M \omega r'r''}{i \hbar \sin \omega T} \right)
\;   ,
\label{eq:propagator_RHO}
\end{equation}
with a conformal parameter index given by Eq.~(\ref{eq:ang-momentum_ISP-extension}).
The path-integral result~(\ref{eq:propagator_RHO}) for the ``radial harmonic oscillator'' was
first derived by Peak and Inomata~\cite{Peak-Inomata:1969}, and has a broad range of applications; 
it is established by using Weber's second exponential integral for Bessel functions~\cite{wat:44} combined with appropriate recursion relations~\cite{Peak-Inomata:1969,Grosche-PI, Grosche-Steiner_PI-QM, Kleinert-PI}.
 This is briefly reviewed and discussed in Appendix~\ref{app:PI-setup}.
 Incidentally, while Eq.~(\ref{eq:propagator_RHO}) refers to the problem with an inverse square potential,
 i.e., of the form~(\ref{eq:propagator_with-ISP}), we will keep the notation without the tilde for the sake of simplicity.

We will apply the path integral~(\ref{eq:propagator_RHO}) to the characterization of the spectral properties
of the generator $G$ of Eqs.~(\ref{eq:G-as-Hamiltonian})--(\ref{eq:G-as-Hamiltonian_tilde}),
where $\tilde{H}_{G}$, given in Eq.~(\ref{eq:G-as-Hamiltonian_tilde}), is equivalent to
 Eqs.~(\ref{eq:potential_HO+ISP})--(\ref{eq:squared-frequency}).
 This applies to the operator $H$ as well as to $R$ and $S'=-S$, defined in Eqs.~(\ref{eq:operator-R}) and (\ref{eq:operator-S}),
and related to the generic Hamiltonian of Eq.~(\ref{eq:potential_HO+ISP}) with the assignments of 
Eqs.~(\ref{eq:operator-R_parameters}) and (\ref{eq:operator-S_parameters}).
Moreover, from the analysis of Eqs.~(\ref{eq:elliptic-generators})--(\ref{eq:hyperbolic-generators}),
these assignments allow for a thorough characterization of the three operator classes: parabolic, elliptic, and hyperbolic.
In this generic framework, the CQM frequency $\omega$ to use in Eq.~(\ref{eq:propagator_RHO})
is given through Eq.~(\ref{eq:squared-frequency}).

Additionally, for some of the applications below, we will compute the path integrals 
with appropriate analytic continuations.
As will be further analyzed in Sec.~\ref{sec:spectral_CQM_hyperbolic},
in Eqs.~(\ref{eq:operator-R_parameters})--(\ref{eq:operator-S_parameters}), the operator $S$ 
is obtained from $R$ by an analytic continuation $\omega \rightarrow - i \omega$ that completely changes the nature of the spectrum
(both $\omega \rightarrow \mp i \omega$ are equally valid, 
but the negative sign has operational advantages discussed in Sec.~\ref{sec:spectral_CQM_hyperbolic}).
Another such extension is the commonly used 
continuation to Euclidean time; in this context,
for the propagator $K_{l+\nu} (r'',r';T)$,
 the following formal replacement is made: 
$T \rightarrow -iT$.
For the propagator of Eq.~(\ref{eq:propagator_RHO}), the corresponding Euclidean-time path integral reads
 \begin{equation}
 K^{(RHO)}_{l+\nu} (r'',r';-i T) 
=
\frac{M \omega }{ \hbar \sinh \omega T} 
\,
\sqrt{ r' r''} 
\,
\exp \left[ - \frac{M \omega }{2 \hbar } \left(r'^{2} + r''^{2} \right) \coth \omega T \right]
I_{\mu} \left( \frac{M \omega r'r''}{ \hbar \sinh \omega T} \right)
\;   .
\label{eq:propagator_RHO_Euclidean}
\end{equation}

It should be stressed that the propagators~(\ref{eq:propagator_RHO})--(\ref{eq:propagator_RHO_Euclidean}) 
correspond specifically to the Hamiltonian $\tilde{H}$ of the form of Eq.~(\ref{eq:potential_HO+ISP}); for CQM, 
Eq.~(\ref{eq:G-as-Hamiltonian}) provides the connection with the generic Hamiltonian $H_{G}$ 
associated with the generator $G$ (with possibly an extra sign $\sigma$).

In the next sections, for our general description of the generators $G$, we will use the notations 
$K^{(\text{ell})}_{l+\nu} $,
$K^{(\text{par})}_{l+\nu} $,
and
$K^{(\text{hyp})}_{l+\nu} $
for the generic propagators of the three classes of generators.

\section{Path-Integral Spectral Analysis of the Generic Elliptic ($\mathbf R$-Like) Generators}
\label{sec:spectral_CQM_elliptic}

In this section, we analyze the propagators for generic elliptic generators, including the prototypical 
operator $R$.
The analysis of the spectral decomposition is straightforward for these operators,
as it consists of a purely discrete spectrum.

\subsection{Elliptic Generators: Spectral Properties and Path-Integral Propagator}

 From Eq.~(\ref{eq:operator-R_parameters}), it is clear that 
$\displaystyle R $
 can be viewed as an analog harmonic oscillator with an inverse square potential;
 thus, the propagator is given in Eq.~(\ref{eq:propagator_RHO}).
 Moreover, a generic elliptic operator $G$, with a discriminant $\Delta <0$,  
inherits the basic spectral properties of $R$, according to Eq.~(\ref{eq:elliptic-generators}), i.e.,
 $G 
 = \sigma \sqrt{| \Delta |}  \; R$.
 Thus, it follows that the eigenstates of an elliptic operator $G$ are the same as those of $R$, 
 with a spectrum of eigenvalues $\hbar  \mathfrak{g} $ such that
 \begin{equation}
 \mathfrak{g} 
 = \sigma \sqrt{| \Delta |}  \; \mathfrak{r}
\label{eq:eigen-g_eigen-r}
\; 
\end{equation}
 is rescaled from the spectrum $\hbar \mathfrak{r} $ of $R$. 
 In addition, the functional form of the elliptic-generator propagator
 \begin{equation}
 K^{(\text{ell})}_{l+\nu} (r'',r';T) 
 =  
 K^{(RHO)}_{l+\nu} (r'',r';T) 
 \; ,
 \label{eq:propagator_ell=RHO}
 \end{equation}
 where $K^{(RHO)}_{l+\nu} $ is given in Eq.~(\ref {eq:propagator_RHO}),
  applies to this entire family, with a CQM frequency $\omega = \sqrt{|\Delta|}/2$, according to Eq.~(\ref{eq:squared-frequency}).

\subsection{Elliptic Generators: Spectral Analysis}

The spectral decomposition of the relevant 
operators, in terms of eigenfunctions and eigenvalues,
can be obtained by performing expansions of the propagator through established identities,
with series or integrals that select appropriate discrete or continuous parts of the spectrum.
This can be done by identifying expressions of the form
$f(A )= \sum_{n} f(a_{n}) P_{n} $, and corresponding generalizations with integrals, 
for a given operator $A$ with eigenvalues $a_{n} $, where $f$ is a function of the operator and
$P_{n} = \left| n \right\rangle \left\langle n \right| $ is the orthogonal projector.
The most basic expansion of this kind in the path-integral approach is provided by the propagator itself, due to its basic definition~(\ref{eq:propagator_QM_from_evolution_operator}) in terms of the time evolution operator. 
As shown below, this leads to a discrete spectrum for the elliptic operators.

The spectral decomposition for the operator $R$ and all elliptic generators
 can be made explicit using the bilinear generating function of associated Laguerre polynomials known as the
 Hille-Hardy formula~(pp. 189-190 in Ref.~\cite{Erdelyi_HTF2}),
  \begin{equation}
 \frac{1}{ (xy z)^{\mu/2} ( 1-z )}
\,  \exp \left[ - z \,  \frac{(x+y)}{(1-z )} \right]
 \,
 I_{\mu} \left( \frac{2\sqrt{xy z}}{1-z} \right)
= \sum_{n=0}^\infty 
\frac{ n!}{\Gamma(n+\mu+1)}
\,
L_n^{(\mu)}(x)
L_n^{(\mu)}(y) \, z^{n}
\; ,
\label{eq:Hille-Hardy}
\end{equation}
with $|z|<1$.
The definition of normalization of the generalized Laguerre polynomials used in Eq.~(\ref{eq:Hille-Hardy})
is that of Ref.~\cite{Erdelyi_HTF2}, corresponding to its relation
$\displaystyle L_n^{(\mu)}(x)
= 
\binom{n + \mu}{n} \,
_{1}F_{1} (-n, \mu + 1; x)$  to the confluent hypergeometric function.

Then, the spectral-series expansion of the Euclidean-time propagator $K^{(\text{ell})}_{l+\nu}(r'',r';-iT)$, from Eqs.(\ref{eq:propagator_RHO_Euclidean}) and 
(\ref{eq:propagator_ell=RHO}), 
can be recast in the form  
\begin{equation}
K_{l+\nu}(r'',r';-iT)    
= \sum_n e^{-E_n T/\hbar} \,
\mathcal{U}_{n,l}(r'')
 \, \mathcal{U}_{n,l}^*(r')
\label{eq:spectral-decomposition_series}
\; ,
\end{equation}
with the substitutions 
$x=M\omega{r'}^{2}/\hbar$, 
$y=M\omega{r''}^{2}/\hbar$, 
and
 $ z =e^{-2\omega T}$, so that
$2 z/(1-z) = (1+z)/(1-z) -1= \coth \omega T -1 $.
Equation~(\ref{eq:spectral-decomposition_series}) provides the expansion for the radial propagator in terms of reduced radial wave functions $\mathcal{U}_{n,l}(r)$, 
and a similar expansion involves the 
$d$-dimensional wave functions $ \psi_{n,l,\boldsymbol{m}} ({\bf r})$ 
with the full-fledged propagator of Eq.~(\ref{eq:propagator_partial_wave_exp_Y}).
By direct inspection, 
the eigenvalues $\tilde{E}_{G,n}$ corresponding to $\tilde{H}_{G}$ are of the form
$ \tilde{E}_{G,n}  = \hbar \omega \,
(1 + \mu + 2n)$; thus, from Eqs.~(\ref{eq:eigen-g_E}), (\ref{eq:squared-frequency}), and (\ref{eq:eigen-g_eigen-r}),
and using the eigenvalues $\hbar \mathfrak{r}_{n}$ of $R$,
\begin{equation}
 {E}_{G,n}  =  \sigma \tilde{E}_{G,n} = \hbar \mathfrak{g}_{n} 
 = \sigma \sqrt{| \Delta |}  \; \mathfrak{r}_{n}
 =
\sigma \hbar \omega \,
(1 + \mu + 2n)
 \label{eq:tilde-H_eigenvalues}
\; ,
\end{equation}
where
\begin{equation}
\mathfrak{r}_{n} = \mathfrak{r}_{0} + n 
\; , \; \;   \; \;  
\mathfrak{r}_{0} =\frac{1}{2}  (1 + \mu )
\label{eq:R_eigenvalues}
\; ,
\end{equation} 
and the parameter $\mu $
 is specified by Eq.~(\ref{eq:ang-momentum_ISP-extension});
 in Ref.~\cite{AFF:76}
 this is explicitly written in the form $\mu = 2 \mathfrak{r}_{0} -1$.
 The spectrum~(\ref{eq:R_eigenvalues}) is the discrete series $D_{\mathfrak{r}_{0}}^{\; +}$
bounded below of the unitary, irreducible representations of the group $SO(2,1)$~\cite{wyb:74}.
Furthermore, their common eigenstates 
 $\mathcal{U}_{n,l}(r)$ 
 are identified as
\begin{equation}
 \mathcal{U}_{n,l} (r)=
\sqrt{\frac{2 \, \Gamma(n+1)}{\Gamma( 1 + \mu +n)}}
\,
 r^{-1/2}
 \left( \sqrt{\frac{ M \omega }{\hbar }}  \, r \right)^{\mu + 1}
  \exp \biggl(- \frac{M \omega } {2 \hbar }  \, r^2 \biggr) 
 \, L_n^{(\mu)} \left( \frac{M \omega }{\hbar  } \, r^2 \right)
\label{eq:tilde-H_eigenstates}
\; .
\end{equation}

\subsection{Elliptic Generators: Conclusions}

Some remarks on the nature of the mathematical results for elliptic operators are in order.
The derivation of the spectral properties for elliptic generators involves an expansion of the 
form~(\ref{eq:spectral-decomposition_series}), which relies on the well-established 
identity~(\ref{eq:Hille-Hardy}).
The Euclidean-time framework permits a rigorous procedure for the 
identification of the corresponding eigenstates and eigenvalues.
This is a well-known theorem for the radial harmonic oscillator~\cite{Grosche-PI, Grosche-Steiner_PI-QM, Kleinert-PI},
which corresponds to our problem via the analog representation~(\ref{eq:propagator_ell=RHO}).
 
 These rigorous results are valid for arbitrary dimensionality $d$; in particular, they
 agree with the corresponding expressions in Sec.~4 of the original 
dAFF model of Ref.~\cite{AFF:76} for $d=1$.
It should be noted that, for $d>1$, the value of $\mathfrak{r}_{0}$ is also dependent on the angular momentum $l$.
The $d$-dimensional wave functions $\psi_{n} ({\bf r})$
can be assembled by considering 
the corresponding spectral expansion of the propagator
of Eq.~(\ref{eq:propagator_partial_wave_exp_Y}), whence
\begin{equation}
\psi_{n, l, \boldsymbol{m}
} ({\bf r})  = 
r^{-(d-1)/2} \,  
\mathcal{U}_{n,l} (r)
\, Y_{l \boldsymbol{m}} ({\bf \Omega})
\; ,
\end{equation}
 which only changes the factor $r^{-1/2}$ in Eq.~(\ref{eq:tilde-H_eigenstates})
 to $r^{-d/2}$, and inserts the angular dependence.
As expected, the derived results can be interpreted as a rescaled, ISP-extended version of those for the standard isotropic oscillator in quantum mechanics. 
 For example, the energy levels of the isotropic oscillator are given by 
$E_{n} = \hbar \omega (N+d/2)$, with principal quantum number $N = 2n+ l + \nu +1$ 
(and $n$ being the radial quantum number), which is extended with the
rule~(\ref{eq:ang-momentum_ISP-extension}), so that $\tilde{E}_{n} =
\hbar \omega (2n + \mu + 1)$, as in Eq.~(\ref{eq:tilde-H_eigenvalues}). 
A similar replacement~(\ref{eq:ang-momentum_ISP-extension}) for the wave functions confirms this interpretation for 
Eq.~(\ref{eq:tilde-H_eigenstates}).

\section{Path-Integral Spectral Analysis of the Generic Parabolic ($\mathbf H$-Like) Generators}
\label{sec:spectral_H_Fourier}

In this section, we analyze the propagators for generic parabolic generators, including the prototypical 
operator $H$.
The case of $H$ (and parabolic operators by extension) is particularly interesting because it describes the initial 
Hamiltonian, and a number of approaches can be successfully applied, including a novel technique 
(``Fourier method,'' Appendix~\ref{app:Fourier})
to derive the eigenfunctions of operators with a continuum spectrum.

\subsection{Hamiltonian $H$ and Parabolic Generators: Spectral Properties
and Path-Integral Propagator}
\label{subsec:spectral_H}

A generic parabolic operator $G$, with a discriminant $\Delta =0$, 
inherits the basic spectral properties of the original Hamiltonian $H$, 
 according to Eq.~(\ref{eq:parabolic-generators}), i.e.,
 $G  = \sigma \, H$.
Its eigenstates are the same as those of $H$, 
 with a rescaled spectrum of eigenvalues $\hbar  \mathfrak{g} $ such that
$ 
 \mathfrak{g} 
 = \sigma E/\hbar$ is determined
  from the spectrum $E $ of $H$.
 In particular, the functional form of the propagator
 $K^{(H)}_{l+\nu} $
 applies without any modifications to this family of parabolic generators,
 with a CQM frequency $\omega =0$.
 
Clearly, the 
 Hamiltonian operator $H$ corresponds to a pure inverse square potential. The corresponding path
integral could be computed separately (see Appendix~\ref{app:PI-setup}), but 
it is straightforward to derive it by taking the limit $\omega\rightarrow 0$ of the 
generalized harmonic oscillator propagator of Eq.~(\ref{eq:propagator_RHO}).
Thus, the propagator corresponding to the operator $H$ and other members of its
parabolic family is given by 
\begin{align}
\! \! \! \! \! \! 
 K^{(\text{par})}_{l+\nu} (r'',r';T)
=  K^{(H)}_{l+\nu} (r'',r';T)
&  =
 \lim_{\omega\rightarrow 0} 
 K^{(RHO)}_{l+\nu} (r'',r';T)
 \\
 & =
 \frac{M}{i \hbar T} \sqrt{r'r''} \,
 \exp \left[ \frac{iM }{2 \hbar T}({r'}^{2}+{r''}^{2}) \right]
 \,
 I_{\mu} \left( \frac{ M r'r''}{i \hbar T}\right)
 \; .
 \label{eq:propagator_ISP}
 \end{align}
  
  The spectral decomposition and spectral properties of this Hamiltonian model can be analyzed in several ways. 
  We will consider the following 3 methods: 
(i) the zero-frequency limit of the spectral decomposition of the radial harmonic oscillator,
using the Mehler-Heine formula for generalized Laguerre polynomials;
(ii) direct evaluation of the spectral decomposition from the propagator~(\ref{eq:propagator_RHO}) 
using Weber's second exponential integral;
(iii) computation of the eigenfunctions with the Fourier transform of the propagator~(\ref{eq:propagator_RHO}),
using an integral representation of the product of Bessel functions.

\subsection{Parabolic Generators: Zero-Frequency Limit of the Propagator---Mehler-Heine Formula}
\label{subsec:zero-frequency_Mehler-Heine}

The first method uses the Mehler-Heine limit formula for the Laguerre 
polynomials~(p. 191 in Ref.~\cite{Erdelyi_HTF2}),
\begin{equation}
\lim_{n \rightarrow \infty}
n^{-\mu} L_{n}^{(\mu)} \left( \frac{x}{n} \right)
=
x^{-\mu/2} \, J_{\mu} ( 2 \sqrt{x} )
\; ,
\label{eq:Mehler-Heine}
\end{equation}
with the following sequence of formal steps.
Starting from the spectral decomposition expansion~(\ref{eq:spectral-decomposition_series}),
with
Eqs.~(\ref{eq:tilde-H_eigenvalues}) and (\ref{eq:tilde-H_eigenstates}), and $\sigma = 1$,
we can turn the series into an integral by realizing that 
the eigenvalue level spacing $E_{n+1}-E_{n} = 2 \hbar \omega$ asymptotically approaches zero. 
Thus, the energy values $E_{n}$ approach the zero limit except when $n \rightarrow \infty$, 
so that the actual energy levels $E$ of the continuum 
satisfy the asymptotic approximate replacement $E \approx 2 \hbar \omega n$. 
Correspondingly, the asymptotic conversion into an integral follows the rule
$\sum_{n} \approx \int dE/(2 \hbar \omega)$, which amounts to $\sum_{n} 1/n \approx \int dE/E$.
In addition, defining $x/n = Mr^{2} \omega /\hbar $ to match the argument of the Laguerre polynomials, 
this gives $x=k^2r^2/4$, with $k$ such that $E=  \hbar^{2} k^2/2M$, 
and the argument of the Bessel function in Eq.~(\ref{eq:Mehler-Heine}) becomes 
$2 \sqrt{x} = kr$.
Then, approximating the eigenfunction coefficient in Eq.~(\ref{eq:tilde-H_eigenstates}),
for $n \gg 1$,
as $c_{n,l} = \sqrt{2 \, \Gamma(n+1)/\Gamma( 1 + \mu +n)}
\approx \sqrt{2/n^{\mu}}$, 
and $e^{-M\omega r^2/2 \hbar} \approx 1$,
the asymptotic discrete eigenfunctions become
\begin{displaymath}
\mathcal{U}_{n,l} (r)
 \approx \sqrt{2} \, n^{-1/2} (x/r)^{1/2} J_{\mu} (kr)
\; ,
\end{displaymath}
where $x/r= k^2 r/4$.
Thus, Eq.~(\ref{eq:spectral-decomposition_series}),
for $n \gg 1$, turns into
\begin{align} 
K^{(\text{par})}_{l+\nu} (r'',r';T)
&
 \approx
\int_{0}^{\infty}
  \frac{dE}{E} \, n \, e^{-iET/\hbar} \,
\mathcal{U}_{n,l} (r'') \, \mathcal{U}^{*}_{n,l} (r') 
\nonumber
\\
&
\approx 
\int_{0}^{\infty}
 dE e^{-iET/\hbar} 
\, \left[ \sqrt{\frac{M}{\hbar^2}} \, \sqrt{r} J_{\mu} (kr'') \right]
\left[ \sqrt{\frac{M}{\hbar^2}} \, \sqrt{r} J_{\mu} (kr') \right]
 \label{eq:propagator_ISP_spectral-decomposition_asymp} 
\; , 
\end{align}
which takes the form of the continuous spectral decomposition
\begin{equation}
K^{(\text{par})}_{l+\nu} (r'',r';T)
=
 \int_{0}^{\infty}
  dE e^{-iET/\hbar} \, 
  \mathcal{U}_{E,l} (r'') \, \mathcal{U}^{*}_{E,l} (r')
\; ,
 \label{eq:propagator_ISP_spectral-decomposition}
  \end{equation}
where the continuous energy eigenfunctions are given by
  \begin{equation}
  \mathcal{U}_{E,l}(r) 
  = \sqrt{\frac{M}{\hbar^2}} \, \sqrt{r} J_{\mu} (kr)
  \; ,
  \label{eq:ISP_energy-eigenfunctions}
  \end{equation}
with $k = \sqrt{2ME/\hbar^2}$.
By the normalized form of  the expansion~(\ref{eq:propagator_ISP_spectral-decomposition}),
the eigenstates~(\ref{eq:ISP_energy-eigenfunctions}) are Dirac-normalized with respect to the variable $E$.
Therefore, through the zero-frequency limit, we have derived the continuous analog of the spectral decomposition 
series~(\ref{eq:spectral-decomposition_series}), 
for a spectrum with values restricted to the half-energy line $E \in [0, \infty)$
and eigenstates~(\ref{eq:ISP_energy-eigenfunctions}).

\subsection{Parabolic Generators: Spectral Decomposition via Weber's Second Exponential Integral}
\label{subsec:explicit-continuous_Weber}

The second method involves writing the spectral decomposition directly 
in its continuous spectral form~(\ref{eq:propagator_ISP_spectral-decomposition}), which
will be expressed in the Euclidean-time version,
\begin{equation}
 K^{(\text{par})}_{l+\nu} (r'',r';-iT)
 = \int_{0}^{\infty}
  dE \,
 e^{-ET/\hbar} \,
   \mathcal{U}_{E,l} (r'') \, \mathcal{U}^{*}_{E,l} (r')
 \; ,
 \label{eq:propagator_ISP_spectral-decomposition_E} 
\end{equation}
where $K^{(\text{par})}_{l+\nu} $ is given by Eq.~(\ref{eq:propagator_ISP}),
and the corresponding real-time propagator involves the replacement $T \rightarrow iT$.
No additional assumptions are needed as the propagator is expanded in this form by construction
from a well-established integral identity 
for Bessel functions~(Sec.~13.31 of Ref.~\cite{wat:44} and 10.22.67 of Ref.~\cite{NIST:2010}):
 Weber's second exponential integral~(\ref{eq:Weber-2nd-exponential-int}).
With the substitutions $E= x^2$, $T/\hbar =  c^2$,
 $\sqrt{2 M} \, r' /\hbar=a $, $\sqrt{2M} \, r''/\hbar =b$
  in Eq.~(\ref{eq:Weber-2nd-exponential-int}), and insertion of an additional factor $2M\sqrt{r'r''}/\hbar^2$,
  the identity turns formally into the desired Eq.~(\ref{eq:propagator_ISP_spectral-decomposition_E}),
  with the same continuous energy eigenfunctions as displayed in Eq.~(\ref{eq:ISP_energy-eigenfunctions}).

\subsection{Parabolic Generators: Spectral Decomposition via a Fourier Method}
\label{subsec:Fourier-method}

The third method is based on a simple restatement of the spectral expansion of an operator with a continuous spectrum.
From the continuous spectral decomposition~(\ref{eq:propagator_ISP_spectral-decomposition}),
the inverse Fourier transform with respect to the variable $E$ converts the propagator 
$ K^{(\text{par})}_{l+\nu} (r'',r';T)$ into the wave-function product 
\begin{equation}
  \mathcal{U}_{E,l} (r'') \, \mathcal{U}^{*}_{E,l} (r')
=
\frac{1}{2 \pi \hbar} \,
\int_{-\infty}^{\infty} dT \, \exp \left( \frac{iET}{\hbar} \right) \, K^{(\text{par})}_{l+\nu} (r'',r';T)
\; .
\label{eq:eigenfunctions_from-Fourier_ISP}
\end{equation}
This yields an explicit Fourier-integral representation of the spectral-decomposition wave-function product.
Evaluation of this spectral integral can be established 
from the integral representation~(\ref{eq:Bessel-product-integral}) of the product of Bessel functions
(Sec.~13.7
 of Ref.~\cite{wat:44}).
 In practice, we are interested in the limit 
 $c \rightarrow 0^{+}$, for which the integral~(\ref{eq:Bessel-product-integral}), with $s=c + 2 i t$, becomes
 \begin{equation}
J_{\mu} (z')  J_{\mu} (z'') 
= \frac{1}{2\pi  i}
\int_{-\infty}^{\infty}
\, \exp \left( it \right) \,
\exp \left[  i \frac{ ( z'^{\, 2} + z''^{\, 2}  ) }{4  t } \right] 
\, I_{\mu} \left( \frac{ z' z'' }{2 i  t } \right) \frac{d t}{ t }
\; .
\label{eq:Bessel-product-integral-2}
\end{equation}
 In Eq.~(\ref{eq:Bessel-product-integral-2}),
 the following substitutions are made on the right-hand side:
$z'= k r'$, $z'' = k r''$, and
$t =  ET/\hbar$, with the time interval $T$ as integration variable. Then, 
the exponential factors become 
$\displaystyle \exp [ {i(z'^2+z''^2)/4t} ] = \exp [ {iM( r'^{2}+r''^{2} )/(2\hbar T)} ] $, 
with the argument of the Bessel function being $z' z''/2it  = M r'r''/i\hbar T$.
This shows that the integrand of Eq.~(\ref{eq:Bessel-product-integral}) is proportional to the propagator
$K^{(\text{par})}_{l+\nu} $ of Eq.~(\ref{eq:propagator_ISP}) after 
 inserting an additional factor $M\sqrt{r'r''}/\hbar^2 $, so that
\begin{equation}
 \left[ \sqrt{\frac{M}{\hbar^2}} \, \sqrt{r} J_{\mu} (kr'') \right]
\left[ \sqrt{\frac{M}{\hbar^2}} \, \sqrt{r} J_{\mu} (kr') \right]
=
\frac{1}{2 \pi \hbar} \, \int_{-\infty}^{\infty} dT \, \exp \left( \frac{iET}{\hbar} \right) \, K^{(\text{par})}_{l+\nu} (r',r'';T)
\; .
\label{eq:eigenfunctions_from-Fourier_ISP_explicit}
\end{equation}
Comparison of Eqs.~(\ref{eq:eigenfunctions_from-Fourier_ISP}) and (\ref{eq:eigenfunctions_from-Fourier_ISP_explicit}) 
shows that the energy eigenfunctions are given by Eq.~(\ref{eq:ISP_energy-eigenfunctions}).

Incidentally, the explicit form of the wave function product in Eq.~(\ref{eq:eigenfunctions_from-Fourier_ISP})
 can also be established via the zero-frequency limit of the radial harmonic oscillator propagator.
In this limit, the generalized Laguerre polynomials turn into Bessel functions,
and the coefficients of the Fourier series~(\ref{eq:spectral-decomposition_series}) 
(where $E_{n}$ is linear in $n$) turn into the required Fourier integral.
 
 \subsection{Parabolic Generators: Conclusions}

 In conclusion, 
 we have shown by several techniques that the Dirac-normalized continuous 
 eigenstates are given by Eq.~(\ref{eq:ISP_energy-eigenfunctions}).
All the methods yield the same result, with a purely continuous 
spectrum restricted to the non-negative energy half-line $E \in [0, \infty)$.
This result 
is valid for arbitrary dimensionality $d$; in particular, it
agrees with the value in Eq.~(A.18) of the original dAFF model of Ref.~\cite{AFF:76} for $d=1$, 
and can be interpreted with the physical insights known for analog elementary problems (e.g., reduces to the free
particle when $g=0$). For the $d$-dimensional model, the radial wave function is $\mathcal{U}_{E,l}(r)/r^{(d-1)/2}$.

A number of final remarks on the nature of the mathematical results for parabolic operators are in order.
A rigorous characterization of the spectral decomposition 
as a theorem is provided by the second method, which relies on the
 direct use of the well-known Weber's second exponential integral; 
 this is similar to the procedure followed for elliptic operators via the Hille-Hardy formula---however, 
such constructions require an ad hoc guessing of the correct expression that mimics this resolution. 
By contrast, the third method inverts the process and introduces a systematic
calculational procedure whereby the wave functions are obtained by a direct computation, 
provided that the integrals can be performed. 
In Appendix~\ref{app:Fourier}, we elaborate on this novel Fourier method, stating it 
in more general terms and addressing its basic properties; 
in its more general form, it will be used in the next section for hyperbolic operators.
It should also be noticed that
the first method was presented as a formal procedure that shows consistency 
with the more general result for elliptic operators
via an appropriate limit. While these operators, elliptic and parabolic,
are of a different nature, they are still related by a continuous transformation with the parameter $\omega$
 enforcing the asymptotic transition.

Interestingly, the derivations of this section show that there are simple connections between
the known mathematical theorems used.
Specifically, Weber's second exponential integral has the following two properties:
(i) it is a restatement of the continuous limit of the Hille-Hardy formula;
(ii) it can be interpreted as a Fourier transform of the integral representation~(\ref{eq:Bessel-product-integral}).
To our knowledge, these statements are not spelled out in the mathematics literature.

\section{Path-Integral Spectral Analysis of
the Generic Hyperbolic ($\mathbf S$-Like) Generators}
\label{sec:spectral_CQM_hyperbolic}

In this section, we analyze the propagators for generic hyperbolic generators, including the prototypical 
operator $S'$.
The analysis of the spectral decomposition is analytically more involved 
due to the nature of the continuous spectrum.

\subsection{Hyperbolic Generators: Spectral Properties and Path-Integral Propagator}
\label{subsec:hyperbolic_PI}

A generic hyperbolic operator $G$, with a discriminant $\Delta >0$,
 inherits the basic spectral properties of $S$ or $S'$, according to Eq.~(\ref{eq:hyperbolic-generators}), i.e.,
 $G 
 = \sigma \sqrt{\Delta }  \; S'$.
 Thus, we can proceed
 as for the other families of operators:
 the eigenstates of a hyperbolic operator $G$ are the same as those of $S'$, 
 with a spectrum of eigenvalues $\hbar  \mathfrak{g} $ such that
 \begin{equation}
 \mathfrak{g} 
 = \sigma \, \sqrt{ \Delta }  \; \mathfrak{s}'
\label{eq:eigen-g_eigen-s'}
\; 
\end{equation}
is rescaled from the spectrum $\hbar \mathfrak{s}' $ of $S'$.

The behavior of the hyperbolic generators falls within the analog model of the generic extended radial harmonic oscillator Hamiltonian 
$\tilde{H}$ of Eq.~(\ref{eq:potential_HO+ISP}).
 Therefore, the functional form of the propagator
 $K^{(\text{hyp})}_{l+\nu}$ 
 applies to the whole family of hyperbolic generators,
 with a CQM frequency $\omega = -i \sqrt{ \Delta }/2$, from Eq.~(\ref{eq:squared-frequency}).
 The imaginary frequency can be obtained with the prescription to analytically continue the potential according to 
$\omega\rightarrow - i\omega$,
with $\omega$ a real ``frequency.''
The result of the analytic continuation 
$\omega\rightarrow- i\omega$
is to generate an inverted harmonic oscillator that overlaps with the inverse
square potential. By the form of this potential, we can predict that all values of the ``energy''
are possible, forming a continuum from minus infinity to plus infinity.
This physical statement will be verified with the path-integral calculation that follows.
It should be noted that systems with such a continuous spectrum, unbounded from below and above, 
are not common and represent idealized models (as actual physical systems exhibit energy bounds). 
But the treatment of such systems with a general path-integral technique is of theoretical and practical interest,
and the results of the previous subsection can be applied directly. For example, the operator $S'$ (being an
inverted radial oscillator) is related to the one-dimensional inverted harmonic oscillator, which is a system
that is related to a variety of physical problems of current relevance~\cite{IHO_Barton, subramanyan_IHO}.

By comparison with the propagator equation~(\ref{eq:propagator_RHO}), 
which is completely general for complex values of the parameter $\omega$,
and with the replacement $\omega\rightarrow - i\omega$,
 the propagator associated with $S'$ and all hyperbolic operators
 is given by
\begin{equation}
  K^{(\text{hyp})}_{l+\nu} (r'',r';T)=
  \frac{M \omega}{i  \hbar \sinh(\omega T)} \sqrt{r'r''}
\,  \exp \left[  \frac{ i M \omega}{2 \hbar}  \left( {r'}^{2}+{r''}^{2} \right)  \coth (\omega T) \right]
\,
I_{\mu}  \left(\frac{ M \omega r'r''}{i \hbar \sinh(\omega T)} \right)  
\; .
\label{eq:propagator_inverted-RHO}
  \end{equation}
  
Incidentally, either extension
$\omega\rightarrow \pm i\omega$ is equally acceptable. This can be seen directly from the Hamiltonian, or explicitly from the propagator.  However, the choice $\omega\rightarrow- i\omega$
has the operational advantage that it preserves the boundary condition at infinity
corresponding to retarded Green's functions solutions (either bound states or outgoing waves). 
This statement can be verified in an explorative manner by a simple WKB evaluation, which is asymptotically exact and yields a solution function
$\mathcal{U} \sim r^{-1/2 + 2 \kappa} \exp (- M \omega r^2/2\hbar) 
\stackrel{(\omega \rightarrow - i \omega)}{\xrightarrow{\hspace*{1cm}}}
\mathcal{U} \sim r^{-1/2 + 2 i\kappa} \exp (i M \omega r^2/\hbar)$,
where $\kappa = \tilde{E}_{G}/(2 \hbar \omega)$ (see Appendix~\ref{app:consistency-check_diff-eq-S});
 and it is further confirmed by the exact solution in terms of Whittaker functions (see final results in this subsection). The advantage of this rule is that one can directly extrapolate by analytic continuation the correct solutions, including Green's functions, from one sector of the theory to another (for example, as shown below, from the inverted to the regular harmonic oscillator).

\subsection{Hyperbolic Generators: Spectral Analysis}
\label{subsec:hyperbolic_spectra}

The spectral decomposition can be obtained from the Fourier integral~(\ref{eq:eigenfunctions_from-Fourier}),
which, with the substitution $ \zeta = \omega T$ takes the explicit form
  \begin{equation}
    \begin{aligned} 
    \mathcal{U}_{E,l} (r'') \, \mathcal{U}^{*}_{E,l} (r')
   & = 
    \frac{1}{2 \pi \hbar} \, \int_{-\infty}^{\infty} 
dT \,  \exp \left( \frac{iET}{\hbar} \right) \,    K^{(\text{hyp})}_{l+\nu} (r'',r';T)
\\
&
= \frac{1}{2 \pi}
 \frac{M }{ \hbar^{2}} \sqrt{r'r''}   
\,  \frac{1}{i} \,
\underbrace{ \int_{-\infty}^{\infty} \frac{d \zeta}{  \sinh \zeta } \, \exp \left( 2 i \kappa \zeta \right) \, \exp \left( i \beta \coth \zeta  \right) \,
I_{\mu}  \left(  \frac{ \alpha }{i  \sinh \zeta } \right)  }_{\displaystyle \mathcal{I}}
\; ,
\label{eq:eigenfunctions_from-Fourier_inverted-RHO}
 \end{aligned}
  \end{equation}
where
\begin{equation}
\begin{aligned}
&  \kappa= \frac{ \tilde{E} }{2\hbar \omega} \; , \;  \;  \;  \;  
\beta = \frac{1}{2} \left( x' + x'' \right) \; , \;  \;  \;  \;  
 \alpha =  \sqrt{x' x''}  \; ,
\\
& \text{with} \; \; \; \;
  x' \equiv \check{r}^{'2} =   \frac{M  \omega}{\hbar} \,   {r'}^{2}
\; \; \; \; \text{and} \; \; \; \;
 x'' \equiv \check{r}^{''2} =   \frac{M  \omega}{\hbar} \,   {r''}^{2}
 \; .
 \end{aligned}
\label{eq:dimensionless-parameters-I}
\end{equation}
In all the ensuing equations below, the dimensionless coordinate
 $\check{r}= \sqrt{M\omega/\hbar} \; r$ is used.  

The integral $ \mathcal{I}$ in Eq.~(\ref{eq:eigenfunctions_from-Fourier_inverted-RHO}) can be
evaluated  by using the approach of Ref.~\cite{Buchholz:1969} (Sec.~6.1).
 This involves developing integral representations for the product of two
 Whittaker functions $M_{\kappa,\mu/2}(z)$ and $W_{\kappa,\mu/2}(z)$ 
 (in different combinations of $M$ and $W$ as
 well as function indices);
 the regularized Whittaker function
 $\mathcal{M}_{\kappa,\mu/2}(z)
 = M_{\kappa,\mu/2} (z)/\Gamma (1+\mu)$ 
 of Ref.~\cite{Buchholz:1969} 
 (that removes singular behavior at negative integer values of $\mu$)
 is also used in the equations below.
  In Appendix~\ref{app:Bessel-Whittaker-products}, we briefly review the basics of these functions,
 which are related to the confluent hypergeometric functions $M(a,b,z)$ and $U(a,b,z)$, and 
 consider the operational procedure of Ref.~\cite{Buchholz:1969} that defines the integral representations,
 with miscellaneous substitutions.\
 For the calculation of $ \mathcal{I}$ in Eq.~(\ref{eq:eigenfunctions_from-Fourier_inverted-RHO}), we are 
 going to use two such specific representations.

\subsubsection{Spectral Decomposition of Hyperbolic Operators via the Fourier Method---Whole-Line Integral}
\label{subsubsec:hyperbolic_Fourier-method_whole-line}

 The first computation of  $ \mathcal{I}$ is based on the direct evaluation of the integral 
 for the entire real line according to the novel identity~(\ref{eq:Whittaker-product-int_4}), which implies
\begin{equation}
     \mathcal{I} =
      e ^{ \pi \kappa}
      \frac{ \Gamma \left( \frac{1+\mu }{2} +i \kappa\right)
 \, \Gamma \left( \frac{1+\mu }{2} - i \kappa \right) }{ \left( x' x'' \right)^{1/2} }
     \,
 \mathcal{M}_{- i \kappa,\mu/2}  ( i x'' )
 \, 
  \mathcal{M}_{i \kappa,\mu/2} ( -ix' )   \, 
 \; ,
    \label{eq:Whittaker-product-int_5}
    \end{equation}
 where the limit $c \rightarrow 0$ is taken, as in discussed Appendix~\ref{app:Bessel-Whittaker-products}.
 This approach has the distinct advantage of being direct and straightforward [provided that Eq.~(\ref{eq:Whittaker-product-int_4}) is established].
 Moreover, due to the symmetry of the propagator with respect to an exchange of $r'$ and $r''$
 [thus, an exchange of $x'$ and $x''$, according to Eq.~(\ref{eq:dimensionless-parameters-I})],
 the products can be written in either equivalent form
 \begin{equation}
 \mathcal{M}_{- i \kappa,\mu/2}  ( i x'' )\,
   \mathcal{M}_{i \kappa,\mu/2} ( -ix' )   
=   \mathcal{M}_{i \kappa,\mu/2} ( -ix'' )   \, 
 \mathcal{M}_{- i \kappa,\mu/2}  ( i x' )
 \; ,
    \label{eq:Whittaker-product-symmetry}
    \end{equation}
 which can also be explicitly verified via a double application of 
  the Whittaker identity of Eq.~(\ref{eq:Whittaker-M-analytic-cont}) below.

\subsubsection{Spectral Decomposition of Hyperbolic Operators via the Fourier Method---Half-Line Integrals}
\label{subsubsec:hyperbolic_Fourier-method_half-line}

 There is a second computation of  $ \mathcal{I}$, which uses an alternative (though related) representation from Ref.~\cite{Buchholz:1969}: Eqs.~(\ref{eq:Whittaker-integral_half-line_BU}), and 
 (\ref{eq:Whittaker-integral_half-line_main}),
  where the latter is another novel identity.
 This second approach, while more involved, provides a separate check and additional information,
 showing the different roles played by the positive and negative energy values, 
 as well as their relationship to the Green's functions. 
 Specifically, in 
 Eq.~(\ref{eq:eigenfunctions_from-Fourier_inverted-RHO}), the integral $\mathcal{I}$ is split into the 
 two contributions from the half-axes (of positive and negative times): $\mathcal{I}= \mathcal{I}_{+} + \mathcal{I}_{-}$,
 with $ \mathcal{I}_{+} $ and $ \mathcal{I}_{-} $ 
defined by
   \begin{equation}
  \mathcal{I}_{\pm}
  =
  \int_{L_{\pm}} \frac{d \zeta}{  \sinh \zeta } \, \exp \left( 2 i \kappa \zeta \right) \, \exp \left( i \beta \coth \zeta  \right) \,
I_{\mu}  \left(  \frac{ \alpha }{i  \sinh \zeta } \right) 
  \; ,
\label{eq:Whittaker-integral_half-line_defs}
   \end{equation}
 where $L_{+} =[0, \infty)$ and $L_{-} =(-\infty, 0]$.
 The integrals~(\ref{eq:Whittaker-integral_half-line_defs}) 
 are explicitly functions of the given parameters:
 (i) the indices $\kappa$ and $\mu$; and (ii) the variables $\alpha$ and $\beta$, or alternatively
 $x'$ and $x''$. The reversal of the sign of $\zeta$ in the integrand of $L_{-}$ compared to $L_{+}$ implies 
 the existence of straightforward symmetries with respect to the parameters of the integrand, 
 such that $ \mathcal{I}_{\pm}$ are not independent but related by
   \begin{equation}
 \mathcal{I}_{-}(\kappa, \mu; x', x'')
 =
 -
  \mathcal{I}_{+}(-\kappa, \mu; -x', -x'')
  \; 
\label{eq:I_pm-symmetries}
\end{equation}
 [or, by abuse of notation, also
 $ \mathcal{I}_{-}(\kappa, \mu; \alpha, \beta) = -  \mathcal{I}_{+}(-\kappa, \mu; -\alpha, -\beta) $].
 In Eq.~(\ref{eq:I_pm-symmetries}), as well as all the equations considered in this paper, the 
final expressions should be evaluated with the principal values of the multivalued functions involved.
 With the results of Eq.~(\ref{eq:Whittaker-integral_half-line_main}),
 we get 
    \begin{equation}
 \mathcal{I}_{\pm }
 =  i \,  \frac{ \Gamma_{ \mp} }{\sqrt{ x' x''  }} \,
 W_{\pm i\kappa,\mu/2} ( \mp i x_{>} )     \,   \mathcal{M}_{ \pm i\kappa,\mu/2}   ( \mp i x_{<} )
  \; ,
\label{eq:Whittaker-integral_half-line2}
  \end{equation}
  where 
 $x_{>}$  and $x_{<}$
are the greater and lesser of the set $\{x', x''\}$ respectively, and the 
 shorthand $\Gamma_{\pm} =  \Gamma \left( (1+\mu)/2 \pm i \kappa \right) $ is used.
  When adding the integrals $\mathcal{I} = \mathcal{I_{+}} + \mathcal{I_{-}}$, the following two relations 
  are applied:
  (i) the semi-circuital analytic continuation~\cite{Buchholz:1969}
    \begin{equation}
    \mathcal{M}_{  \lambda,\mu/2}   (  e^{\pm \pi i} z)
    =
    e^{\pm (\mu + 1) \pi i/2}
      \mathcal{M}_{- \lambda,\mu/2}   ( z)
    \; .
    \label{eq:Whittaker-M-analytic-cont}
        \end{equation}
  and (ii) the connection formula~\cite{Buchholz:1969}
       \begin{equation}
       \mathcal{M}_{  \lambda,\mu/2}   ( z)
    =
   e^{\pm \lambda \pi i}
   \,
   \left[
   \frac{1}{\Gamma_{+}}
   \,
   e^{\mp  i (\mu + 1) \pi i/2} \,
   W_{ \lambda,\mu/2} (  z)     
     +
     \frac{1}{\Gamma_{-}}
     \,
       W_{- \lambda,\mu/2} ( e^{\pm \pi i}  z)    
      \right]
    \; .
\label{eq:Whittaker-M-W-connection}
\end{equation}
The Whittaker functions 
$ {\mathcal M}_{\lambda,\mu/2}  (z)$
and
$W_{\lambda,\mu/2} (z)  $
have to be examined to guarantee results valid within their principal
branches, where the branch cut is conventionally taken as the negative real
half-axis (thus, with the arguments $-\pi <\text{arg} (z) \leq \pi$).
For the present calculation, this involves using the lower and upper signs
of the identities~(\ref{eq:Whittaker-M-analytic-cont})
and (\ref{eq:Whittaker-M-W-connection}) (with $\lambda = i \kappa$) respectively, as applied to
$ \mathcal{I}  
=  \mathcal{I}_{+}  +  \mathcal{I}_{+}$ in that order, so that
\begin{equation}
 \! \! \! \! 
    \mathcal{I}  
= i \,  \frac{ \Gamma_{ +} \Gamma_{ -}  }{\sqrt{ x' x''  }} \,
\underbrace{ \left[
 \frac{1}{ \Gamma_{+} }
\,  W_{ i\kappa,\mu/2} (- i x_{>} )
\! \! \! \! \! \! \! \! \! 
 \underbrace{ \mathcal{M}_{ i\kappa,\mu/2} (- i x_{<} ) 
 }_{ \displaystyle e^{-  (\mu + 1) \pi i/2} M_{ -i \kappa, \mu/2} ( i x_{<} )  }
 \! \! \! \! \! \!  \! \! \! \!    \, 
 +
  \frac{1}{ \Gamma_{-} }
 W_{- i\kappa,\mu/2} ( i x_{>} )     \,   \mathcal{M}_{ -i\kappa,\mu/2}   ( i x_{<} )
 \right]
 }_{\displaystyle      e ^{ \pi \kappa} \, \mathcal{M}_{i \kappa,\mu/2} ( -ix_{>} )   \, 
 \mathcal{M}_{- i \kappa,\mu/2}  ( i x_{<} )
 }
  \; ,
  \label{eq:I-from-Whittaker-integral_half-line2}
  \end{equation}
which reduces to the same value as before, 
Eq.~(\ref{eq:Whittaker-product-int_5}), for $x'> x''$;
however, due to the symmetry of Eq.~(\ref{eq:Whittaker-product-symmetry}), this is also true
for $x'< x''$, showing the equality of both results without restriction. 
 
 \subsubsection{Spectral Decomposition of Hyperbolic Operators---Green's Functions}
\label{subsubsec:hyperbolic_Green-functions}

 Finally, the integrals $\mathcal{I}_{\pm}$ in Eq.~(\ref{eq:Whittaker-integral_half-line2}) give the retarded/advanced Green's functions 
\begin{equation}
G^{(\pm)}_{l+\nu} (r'',r';E)
=
  \mp    \frac{i}{ \hbar \omega } 
 \, 
\frac{  \Gamma \bigl( (1+\mu)/2 \mp i\kappa \bigr) }{ \Gamma (1 + \mu )}  \,
\,
 \frac{1}{\sqrt{ r' r''  }} \,
 W_{\pm i\kappa,\mu/2} ( \mp i \check{r}_{>}^{2} )     \,   M_{ \pm i\kappa,\mu/2}   ( \mp i \check{r}_{<}^{2} )
\; , 
\label{eq:GF_RHO}
\end{equation}
where the jump in the Green's functions according to Eq.~(\ref{eq:eigenfunctions_from-Green})
provides another, related proof of Eq.~(\ref{eq:eigenfunctions_from-Fourier_inverted-RHO}). 

Incidentally, an analytic continuation back to real frequencies can be enforced via $\omega \rightarrow i \omega$,
i.e., $i \kappa \rightarrow \kappa$.
Notice the sign reversal compared to the analytic continuation
that gave Eq.~(\ref{eq:propagator_inverted-RHO}).
Then,
the Green's functions
$G^{(\pm)}_{l+\nu} (r'',r';E) =
  \mp     ( \hbar \omega )^{-1}
 \, 
 \Gamma_{\mp} \,
 W_{\pm \kappa,\mu/2} ( \pm  \check{r}_{>}^{2} )   
   \,   \mathcal{M}_{ \pm \kappa,\mu/2}   ( \pm  \check{r}_{<}^{2} )/  \sqrt{ r' r''  }
 $ 
 for the elliptic generators 
 or ordinary radial harmonic oscillator are obtained,
with $\check{r}_{}^{2} $ given in 
 Eq.~(\ref{eq:dimensionless-parameters-I})
 and
 $\Gamma_{\pm} =  \Gamma \left( (1+\mu)/2 \pm  \kappa \right) $.
 This is a familiar result~\cite{Kleinert-PI}
that also allows a rederivation of their spectrum, Eqs.~(\ref{eq:tilde-H_eigenvalues})--(\ref{eq:R_eigenvalues}).
 The critical difference in the spectral behaviors of hyperbolic generators (continuous operators) versus 
 elliptic generators (discrete operators) arises from the asymptotic behavior of 
 $M_{\pm i\kappa,\mu/2} ( \mp i \check{r}^{2} ) $ versus
 $M_{\pm \kappa,\mu/2} ( \pm \check{r}^{2} ) $ (both with $\kappa \in \mathbb{R} $),
 when $r \rightarrow \infty$, which forces the reduction of the latter to generalized Laguerre polynomials, with
 a discrete quantization of the spectrum.

\subsection{Hyperbolic Operators: Conclusions}
\label{subsec:hyperbolic_eigenvalues-eigenfunctions}

 As a consequence of Eqs.~(\ref{eq:eigenfunctions_from-Fourier_inverted-RHO}),
 (\ref{eq:Whittaker-product-int_5}) [or Eq.~(\ref{eq:I-from-Whittaker-integral_half-line2})],
  and (\ref{eq:Whittaker-product-symmetry}), 
the final result of this calculation is the wave function product
 \begin{align}
 \! \! \!
  \mathcal{U}_{E,l} (r'') \, \mathcal{U}^{*}_{E,l} (r')
  & =
    \frac{1}{ 2 \pi \hbar \omega } \,
     \;
    \Gamma_{-}  \Gamma_{+} 
      \, e^{\pi\kappa}     \,
 \frac{ \mathcal{M}_{i\kappa,\mu/2} ( -i \check{r}^{''2}) 
\, \mathcal{M}_{-i\kappa,\mu/2} ( i \check{r}^{'2}) }{\sqrt{r'' r'} }
 \label{eq:eigenfunctions_from-Fourier_inverted-RHO_explicit-1}
  \\
  & =
  \frac{1}{ 2 \pi \hbar \omega } \,
     \; \Gamma_{-}  \Gamma_{+} 
 \, e^{\pi\kappa}     \,
 \frac{ \mathcal{M}_{-i\kappa,\mu/2} (i \check{r}^{''2}) 
\, \mathcal{M}_{i\kappa,\mu/2} ( -i \check{r}^{'2}) }{\sqrt{r'' r'} }
\; .
\label{eq:eigenfunctions_from-Fourier_inverted-RHO_explicit-2}
 \end{align}
 Again, the equivalence of Eqs.~(\ref{eq:eigenfunctions_from-Fourier_inverted-RHO_explicit-1})
and (\ref{eq:eigenfunctions_from-Fourier_inverted-RHO_explicit-2})
is due to the symmetry of the propagator with respect to the end points, and
 enforced via the analytic continuation of Eq.~(\ref{eq:Whittaker-M-analytic-cont}).
  
Two important conclusions follow from this result.
\begin{enumerate}[(i)]
\item
The spectrum is indeed a continuum, from minus to plus infinity, as 
the solution above applies to all such values of the effective energy 
$\tilde{E}_{G} \equiv \sigma E_{G}$, or [from Eqs.~(\ref{eq:hyperbolic-generators}) and (\ref{eq:dimensionless-parameters-I})]
the parameter 
\begin{equation}
\kappa = \sigma \frac{ E_{G} }{ 2 \hbar \omega } = \sigma \frac{\mathfrak{g}}{\sqrt{ \Delta }} = \mathfrak{s}'  \in (-\infty, \infty)
\label{eq:hyperbolic-Whittaker-index}
\; ,
\end{equation}
which is a dimensionless value covering the whole real axis and equal to the first Whittaker index $\kappa$,
and such that $\mathfrak{s}' = \kappa$ for the eigenvalues $\hbar \mathfrak{s}'$ of the operator
$\displaystyle S' = -S$.

\item The wave functions can be read off from either one of
 Eqs.~(\ref{eq:eigenfunctions_from-Fourier_inverted-RHO_explicit-1})
or (\ref{eq:eigenfunctions_from-Fourier_inverted-RHO_explicit-2}),
by comparison with Eq.~(\ref{eq:eigenfunctions_from-Fourier}), and 
 are given by 
 \begin{align}
   \mathcal{U}_{E,l} (r) 
  \equiv \mathcal{U}_{\kappa, \mu} (r) 
&  =
  \frac{ e^{\pi\kappa/2}  }{ \sqrt{ 2 \pi \hbar \omega } }
\, 
\frac{  \Gamma \bigl( (1+\mu)/2 +i\kappa \bigr) }{ \Gamma (1 + \mu )}  \,
\,
  \frac{
 M_{i\kappa,\mu/2} (-i \check{r}^{2}) }{\sqrt{r}} \, e^{i \chi}
 \label{eq:inverted-RHO_energy-eigenfunctions-1}
  \\
  & =
\frac{ e^{\pi\kappa/2}  }{ \sqrt{ 2 \pi \hbar \omega } }
\, 
\frac{  \Gamma \bigl( (1+\mu)/2 -i\kappa \bigr) }{ \Gamma (1 + \mu )}  \,
\,
  \frac{
  M_{-i\kappa,\mu/2} (i \check{r}^2 ) }{\sqrt{r}} \, e^{i \chi'}
  \; ,
\label{eq:inverted-RHO_energy-eigenfunctions-2}
   \end{align}
   where we have reverted back to the ordinary Whittaker functions
  $ M_{\pm i\kappa,\mu/2} $, and
   $\chi$ and $\chi'$ are arbitrary phase factors.
   The simplest choices are arguably $\chi =0$ or $\chi' =0$,
  but these values are undetermined by the nature of the 
   products~(\ref{eq:eigenfunctions_from-Fourier_inverted-RHO_explicit-1})--(\ref{eq:eigenfunctions_from-Fourier_inverted-RHO_explicit-2}); 
   in particular, the choice of gamma function factors is also arbitrary, as $\Gamma_{\pm} = \Gamma_{\mp}^{*}$.
It should be noted that, in the derivation above,
$ \mathcal{M}_{ -i\kappa,\mu/2 } ( i\check{r}^{2}) \propto
  \mathcal{M}_{ i\kappa,\mu/2 } (-i \check{r}^{2})
$, according to Eq.~(\ref{eq:Whittaker-M-analytic-cont}); thus,
 up to a phase factor, the Whittaker functions 
in Eq.~(\ref{eq:inverted-RHO_energy-eigenfunctions-1})--(\ref{eq:inverted-RHO_energy-eigenfunctions-2}) 
can take either equivalent form
$ M_{ \mp i\kappa,\mu/2 } (\pm  i \check{r}^{2}) $.
\end{enumerate}

 Moreover, the basic results of
 Eqs.~(\ref{eq:eigenfunctions_from-Fourier_inverted-RHO_explicit-1})--(\ref{eq:eigenfunctions_from-Fourier_inverted-RHO_explicit-2}) 
 and~(\ref{eq:inverted-RHO_energy-eigenfunctions-1})--(\ref{eq:inverted-RHO_energy-eigenfunctions-2}) 
are also verified by the analysis of Appendix~\ref{app:consistency-check_diff-eq-S}, 
 based on the associated differential equation.
 
 The following remarks on the nature of the mathematical results for hyperbolic operators are in order.
 First, these results have been established by a rigorous approach that relies on established 
integral representations of the product of Whittaker functions. 
Second, the proof relies on the novel Fourier method for
continuous-spectrum operators presented in Appendix~\ref{app:Fourier}.
Third, they
provide a remarkable example of a continuous spectrum unbounded from both below and above;
and finally, perhaps due to their unusual nature, they have not been addressed in the existing literature:
 ideally, our paper should contribute to fill that gap.

 This concludes our detailed analysis of the most relevant spectral properties of the symmetry operators
 of CQM, with the added bonus of having established appropriate analytic 
 continuation techniques to compare the different types.
 
\section{Conclusions}
\label{sec:conclusions} 

In this work, we have derived a comprehensive path-integral treatment of the symmetry generators of
conformal quantum mechanics (CQM).
This analysis is relevant in the context of CQM as a one-dimensional conformal 
field 
theory~\cite{Jackiw_CFT1-1,Jackiw_CFT1-2,CQM_Okazaki-2015,CQM_Okazaki-2017,Pinzul_CFT1-2017,Khodaee_CFT1-2017,deAlmeida_CFT1-2019,CQM-CFT1_Ardon-2021}.
Moreover, it is noteworthy that the symmetry operators in the weak-coupling regime of CQM are of current interest in understanding the nature of spacetime causal structure in the context of causal diamonds and thermal properties of the vacuum~\cite{Arzano-1,Arzano-2,martinetti-1, martinetti-2, su-ralph-1, su-ralph-2, light-cone, jacobson,Houston_OQS-diamond,RCKF}.

The main results of our analysis include a complete characterization via path integrals of all
the operators from the three families (elliptic, parabolic, and hyperbolic), with distinctly
different spectral and time-evolution properties. In addition to establishing appropriate analytic continuations 
of the path integral for the analog system of a radial harmonic oscillator, 
we have derived novel expressions and a simple general technique
to deal with the spectral characterization of continuous operators.
These results can be used to provide further insight into the physical and mathematical properties of CQM, 
including, inter alia, the physical meaning of this form of SO(2,1) conformal symmetry for near-horizon physics~\cite{Cam:BHT,CQM_MorettiPinamonti-2002}
and applications in quantum cosmology~\cite{Qcosm_Pioline-Waldron-2003, Qcosm_Achour-Livine-2019a,Qcosm_Achour-Livine-2019b}.
Moreover, the general methodology and spectral properties we established for hyperbolic operators are of potential interest in related applications of the 
inverted harmonic oscillator~\cite{IHO_Barton, subramanyan_IHO}, which we are currently investigating, with impact on problems from the quantum Hall effect to black holes, including issues of thermality and complexity~\cite{dalui2020horizon, dalui2020near,bhattacharyya2021multi,qu2022chaos}.

 Another line of work for which the CQM generators, as considered in this paper, are relevant is the recent development of 
Schwarzian mechanics~\cite{SM_Mertens-2017,SM_Lam-2018,SM_Galajinsky-2018,SM_Galajinsky-2019,SM_Fiyukov-2021,SM-dAFF_Masterov-2021}, which is related to the dAFF model. 
Most importantly, Schwarzian mechanics is relevant to the low-energy limit of the 
Sachdev-Ye-Kitaev model~\cite{SYK_Maldacena-2016}, as was
shown in Refs.~\cite{SM_Mertens-2017,SM_Lam-2018}.
 
 Moreover, the use of the symmetry generators $G$ as alternative Hamiltonians with transformed times $\tau$
is still an open question, though some special cases have been suggested in the literature, for example, for the magnetic monopole~\cite{Jackiw:80} and the magnetic vortex~\cite{Jackiw:90}. In Ref.~\cite{CQM_Tada-2018}, the roles played
by the generators $H$, $R$, and $S'$ is considered within a CFT-inspired sine-square deformation, with an interpretation that can be further reexamined within our generalized framework. In addition, the Niederer-Takagi time transformation~\cite{HO-Schr_Niederer-1973, time-transf_Takagi-1990, time-transf_Takagi-1991a, time-transf_Takagi-1991b} involved in the elliptic generators has been used in the physics of cold atoms~\cite{Castin_2004,Wamba_2016}.
However, the interpretation of the time variable is a subtle notion that finds a more natural context in general-relativistic applications, as has been suggested for matrix models~\cite{Strominger_CQM-matrix-2004}. This interpretation has been 
partly implemented for the dynamics of harmonic-oscillator-type Unruh-DeWitt detectors 
in curved spacetimes~\cite{UDW-detect_Hotta_2020}. More generally, this is an issue of relevance in the near-horizon version of CQM for black hole thermodynamics~\cite{Cam:BHT} and causal 
diamonds~\cite{Arzano-1,Arzano-2,martinetti-1, martinetti-2, su-ralph-1, su-ralph-2, light-cone, jacobson,Houston_OQS-diamond,RCKF},
which deserves further investigation.

Finally, the current presentation has been limited to the weak-coupling regime of CQM, as described in Sec.~\ref{sec:intro}. An extension of this framework to the strong coupling regime, with additional implications for renormalization and quantum anomalies, is in progress, and will be reported elsewhere.
\acknowledgments{}
This material is based upon work supported by the Air Force Office of Scientific
Research under Grant No. FA9550-21-1-0017 (C.R.O., A.C., and P.L.D.).
C.R.O. was partially supported by the Army Research Office (ARO), grant W911NF-23-1-0202.
H.E.C. acknowledges support
by the University of San Francisco Faculty Development Fund. 

\bigskip
This article may be downloaded for personal use only. Any other use requires prior permission of the author and AIP Publishing. This article appeared in H. E. Camblong, A. Chakraborty, P. Lopez Duque, C. R. Ordóñez; Spectral properties of the symmetry generators of conformal quantum mechanics: A path-integral approach. J. Math. Phys. 1 September 2023; 64 (9): 092302, and may be found at \url{https://doi.org/10.1063/5.0150349}.
\bigskip

\noindent
{\bf AUTHOR DECLARATIONS}

\noindent
{\bf Conflict of Interest}

\noindent
The authors have no conflicts to disclose.

\smallskip

\noindent
{\bf Author Contributions}

\noindent
All authors contributed equally to this work.

\bigskip

\noindent
{\bf DATA AVAILABILITY}

\noindent
Data sharing is not applicable to this article as no new data were created or analyzed in this study.

\bigskip

\appendix

\section{A Dimensional Form of the Theory of Symmetry Generators and Their Effective Hamiltonian Representation}
\label{app:generators-Hamiltonian_dimensional-rep}

The framework of Subsec.~\ref{subsec:conformal-generator-G} 
is based on the generators $G$ leading to the effective Hamiltonian~(\ref{eq:G-as-Hamiltonian}).
It has the definite advantage of being completely general, but the
Hamiltonian involves canonical variables and time with reduced dimensions, which
are different from those of the original CQM Hamiltonian~(\ref{eq:CQM-general-Hamiltonian})[
(for example, $\tau$ is dimensionless and $[\mathbf{r}]= [\mathbf{Q}]/[f]^{1/2}$, with $[f]=$ time).
While this poses no serious technical challenges, it may be desirable to develop an alternative method 
where the relevant variables have their ``usual'' dimensions, meaning those of the original Hamiltonian $H$.
Specifically, dimensions can be restored to those of $H$ via a characteristic time scale; here,
 this is naturally provided by the factor $u$ of the generalized generator, Eq.~(\ref{eq:gen-generator}), if $u \neq 0$.

In principle, under the assumption $u \neq 0$, 
the rescaling $G= u \hat{H}_{G} (\hat{\mathbf{r}}, \hat{\mathbf{p}} )$, rather than
$G= \sigma \tilde{H}_{G} ({\mathbf{r}}, {\mathbf{p}} )$, 
allows the theory to be redefined consistently in terms of the effective time
$\hat{\tau}  = u \tau $
and canonical variables $\hat{\mathbf{r}},
\hat{ \mathbf{p}} $,
such that $\hat{r}^2= u r^2$ and $\hat{p}^2 = u^{-1}  \mathbf{p}^{2}$.
Then, the reduced effective Hamiltonian becomes
\begin{equation}
 \frac{1}{u}  G \equiv \hat{H}_{G} (\hat{\mathbf{r}}, \hat{\mathbf{p}} )
=
\frac{\hat{p}^2}{2M}
+  \frac{\hbar^2}{2M} \, \frac{g}{\hat{r}^{2}} + \frac{M}{2} \left(-\frac{\Delta}{4 u^{2} } \right) \hat{r}^{2}
\; 
\label{eq:G-as-Hamiltonian_reduced} 
\end{equation}
(with the additional assignments $\lambda = \hbar^2 \, g/M$ and $\hat{p}^{2} = \hat{\mathbf{p}}^{2}$ to be used below).
The evolution operator~(\ref{eq:tau-evolution-op}) admits the alternative form
$U_{\hat{H}}(T) = e^{-i\hat{H} ( \hat{\tau}-\hat{\tau}_{0})/\hbar}$, and the 
CQM frequency becomes
\begin{equation}
\hat{\omega}^{2} = -\frac{\Delta}{4 u^{2} }
\; .
\label{eq:squared-frequency_reduced} 
\end{equation}
With these redefinitions, the time $\hat{\tau}$, the Hamiltonian $ \hat{H}_{G} (\hat{\mathbf{r}}, \hat{\mathbf{p}} ) $, and
the frequency in Eq.~(\ref{eq:squared-frequency_reduced}) have the usual dimensions.

In addition, it may prove useful to define the real and positive time-scale parameter 
\begin{equation}
a = 2 \frac{ |u| }{ \sqrt{ |\Delta |} }
\; ,
\label{eq:time-parameter-a} 
\end{equation}
which generalizes the original parameter $a$ of Ref.~\cite{AFF:76}
 in Eqs.~(\ref{eq:operator-R}) and (\ref{eq:operator-S});
then, $\hat{\omega}^{2} = - \text{sgn} (\Delta)/a^{2}$,
which leads to $\hat{\omega}  = 1/a $ for $\Delta < 0$
and $\hat{\omega}  = -i/a $ for $\Delta > 0$ 
(the choice of sign for the latter is discussed in Sec.~\ref{sec:spectral_CQM_hyperbolic}),
with $\hat{\omega} = 0 $ and  $a=\infty$ for $\Delta = 0$. 
As in Subsec.~\ref{subsec:conformal-generator-G}, 
denoting the corresponding parameter dependence in Eq.~(\ref{eq:G-as-Hamiltonian_reduced})
with $\hat{H} \bigl[ M, g, \hat{\omega} \bigr]$
(where the subscript $G$ is removed for simplicity), we can make the assignments
\begin{align}
& R = \frac{a}{2} \, \hat{H} \bigl[ M, g, \hat{\omega}=1/a \bigr]
\label{eq:operator-R_parameters_dimensional}
\; ,
\\
& H = \hat{H} \bigl[ M, g, \hat{\omega}= 0 \bigr] 
 \; ,
\label{eq:operator-H_parameters_dimensional}
\\
& S' \equiv \frac{a}{2} \, \hat{H} \bigl[ M, g, \hat{\omega}=- i/a \bigr]
\label{eq:operator-S_parameters_dimensional}
\; ,
\end{align}
as $u =a/2$ for both $R$ and $S'$.
The corresponding description of the three classes of operators 
follows from $G = u \hat{H} \left[  M, g, \hat{\omega} \right]$, which also implies that
$G=  \text{sgn} (u) \sqrt{| \Delta |}  \, (a/2) \; \hat{H} \left[  M, g, \hat{\omega} \right]
$ for elliptic and hyperbolic operators; consequently,
\begin{align}
G 
& = \text{sgn} (u) \sqrt{| \Delta |}  \; R
\;  \; \; \text{for elliptic generators}
\; ,
\\
G 
& = u \, H
\;   \; \; \text{for parabolic generators }
\; ,
\label{eq:parabolic-generators_dimensional} 
\\
\text{and} \; \; G 
& = \text{sgn} (u) \sqrt{ \Delta }  \; S'
\;  \; \; \text{for hyperbolic generators}
\; .
\label{eq:hyperbolic-generators_dimensional} 
\end{align}
This restricted framework agrees with the more general characterization 
of Eqs.~(\ref{eq:elliptic-generators})--(\ref{eq:hyperbolic-generators}), where $\sigma$ 
is the more general sign, which can be evaluated from $\text{sgn} (u) $ when $\text{sgn} (u) \neq 0$.
Despite being somewhat restricted, this approach has some appealing features, i.e., its 
dimensional-analysis structure, and includes the all-important operators $R$, $H$, and $S'$.

A final remark is in order regarding the restrictions of this particular approach. We have assumed that $u \neq 0$,
which basically modifies the original inverse-square-potential 
Hamiltonian $H$ with the extension~(\ref{eq:G-as-Hamiltonian_reduced}).
This excludes the family of Hamiltonians $H_{G}$ with $u=0$, for which
the discriminant is $\Delta = v^2$, and which encompasses two subclasses: 
(i) hyperbolic operators with $v \neq 0 $, including $D$ by itself, as well as
 linear combinations of $D$ and $K$;
(ii) parabolic operators with $v=0$, which reduce to simply $K$ (up to an arbitrary multiplication constant $w$).
However, we have seen in
Subsec.~\ref{subsec:conformal-generator-G} that these operators can be handled with the more general framework, 
where $K$ and $D$ are described as equivalent to $H$ and $S'$ respectively.

\section{Path-Integral Framework---Summary of Basic Results}
\label{app:PI-setup}

In this appendix, we summarize the main results on path integration needed to understand
and compute the spectral properties of the operators discussed in the main text.
These include the setup of the time-sliced path integral, the transformation to hyperspherical coordinates,
and the evaluation of the propagator for the radial harmonic oscillator.

\subsection{Path-Integral Framework Setup}
\label{app:PI-summary}

The basic setup starts, as in Sec.~\ref{sec:PI-setup_CQM},
with the path-integral expression~(\ref{eq:propagator_QM}) for the propagator, which
 can be evaluated as the limit of a properly
time-sliced integral in Cartesian coordinates,
\begin{equation}
K_{(d)}({\bf r''}, {\bf r'}; t^{''},t^{'}) 
= 
\lim_{N \rightarrow \infty}
\left( \frac{M}{2 \pi i \epsilon \hbar} \right)^{d N/2}
\,
\left[
\prod_{k=1}^{N-1} \int_{ \mathbb{R}^{d} } d^{d} {\bf r}_{k} \right]
\;
e^{i S^{(N)}/\hbar}
\;  .
\label{eq:propagator_QM2}
\end{equation}
As in Subsec.~\ref{subsec:PI_framework}, we use $t$ for the dynamical time and $\hat{H}$ for a generic Hamiltonian.
The Cartesian form of Eq.~(\ref{eq:propagator_QM2})
involves a time lattice $t_{j}= t'+j \epsilon$,
for the time interval $T=t''-t'$ corresponding to the
 end points
${\bf r}_{0} \equiv {\bf r'}$
and 
${\bf r}_{N} \equiv {\bf r''}$;
in this lattice,
$\epsilon=
T/N$, with $j=0, \cdots,N$,
with $t_{0} \equiv t'$ and $t_{N} \equiv t''$,
such that $ {\bf r}_{j} = {\bf r} (t_{j})$.
The  discrete action 
in Eq.~(\ref{eq:propagator_QM2}) 
is  $S^{(N)}=   \sum_{j=0}^{N-1} S^{(N)}_{j}$, with
$S^{(N)}_{j}
= M
( {\bf r}_{j+1}
- {\bf r}_{j}  )^{2}/2\epsilon -
\epsilon V({\bf r}_{j} )
 $, where $V({\bf r} )$ is the 
potential---if assumed to be time-independent, as is the case for the computation of all the CQM operators,
the result of this path integral is only a function of $T$.

The propagator $ K_{(d)}  ( {\bf r}'', {\bf r}' ; T ) $
can be formally rewritten in non-Cartesian coordinate systems.
It is well-known that caution must be exercised when evaluating Eq.~(\ref{eq:propagator_QM})
in non-Cartesian coordinates, as this typically leads to extra terms of order $\hbar^{2}$ in the action. 
These arise when nonlinear transformations are performed, both in quantum 
mechanics~\cite{Grosche:1987,extra-terms-qm_DeWitt, extra-terms-qm_Edwards, extra-terms-qm_Gervais,extra-terms-qm_Apfeldorf}. 
and quantum field theory~\cite{extra_terms_qft}.
In general, coordinate transformations of Eq.~(\ref{eq:propagator_QM2}) can be applied
{\em before\/} taking the continuum limit, but prescriptions for the choice of position values of the potential 
are dictated by operator ordering.

\subsection{Path-Integral Framework---Hyperspherical 
Coordinates and Besselian Path Integrals}
\label{app:hyperspherical-Besselian}

For the important case of hyperspherical polar coordinates~\cite{Cam_DT1,Cam_DT2,Erdelyi_HTF2},
the $d$-dimensional hyperspherical harmonics
$Y_{l \boldsymbol{m}} ({\bf \Omega})$
and Gegenbauer polynomials $C_{l}^{(\nu)}(x)$ provide the necessary framework for separation of variables;
see Sec.~\ref{sec:PI-setup_CQM} and
Eqs.~(\ref{eq:propagator_partial_wave_exp_Y})--(\ref{eq:propagator_partial_wave_exp}) for the partial wave expansions.
Then, an explicit expression for the propagator can be derived in the time-sliced path integral by transforming
Eq.~(\ref{eq:propagator_QM}) into hyperspherical coordinates
before taking the $N \rightarrow \infty$ limit. 
This is especially useful for central potentials: $ V( {\bf r}) = V(r)$.
The critical step in this derivation is
rewriting the elements of the discretized action 
$S^{(N)}_{j}
= M
( {\bf r}_{j+1}
- {\bf r}_{j}  )^{2}/2\epsilon -
\epsilon V( { r}_{j} )
 $ by separating the angles $\cos \psi_{j+1, j} = \hat{{\bf r}}_{j+1} \cdot \hat{{\bf r}}_{j}$ 
 (with $\hat{ \mathbf{r} } = \mathbf{r}/r$) in a radial-angular resolution 
$\displaystyle
\exp \left( i S^{(N)}_{j} /\hbar \right) = 
 \exp \left[ i  \frac{M}{2 \hbar \epsilon } \, \left( r_{j+1}^2+r_{j}^2 \right)   -i \frac{\epsilon V_{j}}{\hbar} \right] 
\, \exp \left({z_{j} \cos \psi_{j+1, j} } \right)$, where
$\displaystyle z_{j}
=
 \frac{M r_{j} r_{j+1}}{ i \epsilon \hbar}
$.
The product of the last factor $ \exp \left({z_{j} \cos \psi_{j+1, j} } \right)$
for all the intervals ($j=0, \ldots N-1$) 
can be evaluated in the angular integrals of the 
time-sliced path integral, Eq.~(\ref{eq:propagator_QM2}), which involves
multiple applications of $d^{d} {\bf r}_{k} = d r_{k} \, r_{k}^{d-1} \, d \Omega_{k} $, with $d$-dimensional solid-angle integrations.
This is done
 using the degenerate form of
Gegenbauer's addition theorem on partial waves
(generalization of the 3D Rayleigh expansion of a plane wave; Sec.~11.5 of Ref.~\cite{wat:44}):
\begin{equation}
e^{iz\cos \psi}
=
(iz/2)^{-\nu} \Gamma (\nu)
\,
 \sum_{l=0}^{\infty} 
(l+\nu)
I_{l+\nu} (iz ) C_{l}^{(\nu)} (\cos \psi )
\; ,
\end{equation}
where $\nu = d/2-1$ and $I_{p}(x)$
 is the modified Bessel function of the first kind and order $p$.
With this expansion and performing the angular integrations, along with 
Eqs.~(\ref{eq:propagator_partial_wave_exp_Y})--(\ref{eq:propagator_partial_wave_exp}) and the orthonormality of hyperspherical harmonics, 
the path integral for the radial propagator in
Eq.~(\ref{eq:propagator_partial_wave_exp}) becomes
\begin{eqnarray}
K_{l+\nu}(r'',r';T)  = 
\lim_{N \rightarrow \infty}
\! \!
&   & 
\! \!
\left( \frac{M}{2 \pi i \epsilon \hbar} \right)^{N/2}
\,
\prod_{k=1}^{N-1}
\left[ 
 \int_{0}^{\infty} d r_{k}
\right]
\;  
w_{l+\nu}^{(N)} [r^{2}] 
\,
\nonumber  \\
&  \times  &
\,
\exp \left\{
\frac{i}{\hbar} 
R^{(N)}  
 \left[ 
 r_{1},
\dots
,
 r_{N-1}
\right]  (r'', r' ; T)  
\right\}
\;   ,
\label{eq:propagator_QM_spherical_coords_expansion}
\end{eqnarray}
where the radial action is
\begin{equation}
R^{(N)}  
 \left[ 
 r_{1},
\dots
 r_{N-1}
\right]  (r'', r' ; T)  
=
\sum_{j=0}^{N-1}
\left[
\frac{  M
\left( r_{j+1}
- r_{j}  \right)^{2}  
}{2\epsilon} 
-
\epsilon V( r_{j} )
\right]
\; .
\end{equation}
In Eq.~(\ref{eq:propagator_QM_spherical_coords_expansion})
a radial functional weight 
\begin{equation}
w_{l+\nu}^{(N)} [r^{2}]
=
\prod_{j=0}^{N-1} 
\left[ 
\sqrt{2 \pi z_{j}} e^{-z_{j}}
I_{l+\nu} (z_{j})
\right]
\; 
\label{eq:Besselian-functional-weight} 
\end{equation}
has been properly
defined with the radial variables appearing through the characteristic 
dimensionless ratio
$\displaystyle z_{j}
=
 \frac{M r_{j} r_{j+1}}{ i \epsilon \hbar}
$.
 Equation~(\ref{eq:propagator_QM_spherical_coords_expansion})
 [with the restriction to the half-line $r(t) \geq 0$]
 admits the formal Besselian path-integral representation~(\ref{eq:propagator_QM_spherical_coords_expansion_continuum}),
 i.e.,
\begin{equation}
K_{l+\nu}(r'',r';T)
=
\int
{\mathcal D}  r(t)  \,
\; 
w_{l+ \nu} [r^{2}]
\exp 
\left\{
\frac{i}{\hbar} 
\int_{t'}^{t''}
dt
\left[
\frac{M}{2} \,
\dot{r}^{2}
-
V(r)
\right]
\right\}
\;   ,
\label{eq:propagator_QM_spherical_coords_expansion_continuum_app}
\end{equation}
with the angular part of the problem contributing through the nontrivial radial functional weight 
$w_{l+\nu} [r^{2}]$.
As a result of this hyperspherical resolution,
 the property known as interdimensional 
dependence~\cite{interdimensional, Fischer:1992} is exhibited by
Eqs.~(\ref{eq:propagator_QM_spherical_coords_expansion})--(\ref{eq:propagator_QM_spherical_coords_expansion_continuum_app}):
$d$ and $l$ appear in the combination $l+\nu$.

The associated radial action $R[r(t)]$ in Eq.~(\ref{eq:propagator_QM_spherical_coords_expansion})
only includes the radial kinetic energy and the interaction potential,
and excludes the centrifugal potential, whose
role is played instead
 by the functional weight.
The connection
with the usual formulation of the classical radial action
can be shown by a formal argument, using the 
asymptotic form of the Bessel function
$\sqrt{2 \pi z } \, e^{-z} I_{\mu} (z) \sim e^{-(\mu^2-1/4)/2z + O(1/z^2)}$ (for $|z| \rightarrow \infty$)
 in Eq.~(\ref{eq:Besselian-functional-weight}), where $z\sim 1/\epsilon$.
 Thus, this leads to the alternative expression 
 \begin{equation}
K_{l+\nu}(r'',r';T)
=
\int
{\mathcal D}  r(t)  \,
\exp \left\{
\frac{i}{\hbar} 
\int_{t'}^{t''}
dt
\left[
\frac{M}{2} \,
\dot{r}^{2}
-
\frac{\hbar^2}{2M} \frac{ (l+\nu)^2}{r^2}
-
V(r)
\right]
\right\}
\;   ,
\label{eq:propagator_QM_spherical_coords_expansion_cf}
\end{equation}
 where the action does include the centrifugal term---this formal procedure has been called asymptotic recombination~\cite{Inomata_PIs-1992}.
 This equivalence of 
 Eqs.~(\ref{eq:propagator_QM_spherical_coords_expansion_continuum_app})
 and
 (\ref{eq:propagator_QM_spherical_coords_expansion_cf})
 can be established rigorously by the important theorem relating angular momentum 
 and inverse square potential 
 terms---see Eqs.~(\ref{eq:ang-momentum_ISP-extension}) and (\ref{eq:propagator_with-ISP}) 
 and associated statement in Subsec.~\ref{subsec:PI_framework}.
 The latter can be inserted in the path integral 
 [as can be seen in Eq.~(\ref{eq:propagator_QM_spherical_coords_expansion_cf})] by absorbing the 
 inverse-square coupling $g$ as part of an effective angular momentum 
 with $  \mu = \sqrt{(l+ \nu)^2 + g}$.
This was originally shown as above in Ref.~\cite{Peak-Inomata:1969} for $d=3$, 
with a rigorous proof in Refs.~\cite{Fischer:1992,Inomata_PIs-1992}.

\subsection{Inverse Square Potential and Radial Harmonic Oscillator}
\label{app:ISP-radialHO}

The generic radial path integral~(\ref{eq:propagator_QM_spherical_coords_expansion})
can be used to evaluate the propagator for the inverse square potential and radial harmonic oscillator, 
by rewriting it in a more convenient form, 
 \begin{equation}
  K_{l+\nu}(r'',r';T)
=
\lim_{N \rightarrow \infty}
\left(\frac{\alpha}{i} \right)^N
\,
\prod_{k=1}^{N-1}
\left[
 \int_{0}^{\infty} d r_{k} r_{k}
\right]
\,
\prod_{j=0}^{N-1}
\left[
I_{l+\nu}(z_{j}) e^{iA_{j}/\hbar}
\right]
\label{eq:propagator_QM_spherical_coords_expansion_2}
\end{equation}
where 
$\alpha =M/(\epsilon \hbar)$ and
$\displaystyle
A_{j} = \frac{M}{2 \epsilon} 
\left(  {r}_{j+1}^{2}
 +
{r}_{j}^{2} \right)
 -
\epsilon V( r_{j})
$.
The limit 
 $ K_{l+\nu}(r'',r';T)
=
\lim_{N \rightarrow \infty} K_{l+\nu}^{(N)}(r'',r';T)$
in Eq.~(\ref{eq:propagator_QM_spherical_coords_expansion_2})
can be taken by evaluating the $N$-th order term $K_{l+\nu}^{(N)}(r'',r';T)$ 
recursively~\cite{Grosche-PI, Grosche-Steiner_PI-QM, Kleinert-PI, Peak-Inomata:1969}.
 Transforming the path integral into its Euclidean-time form
 (with the replacements $\epsilon =T/N \rightarrow - i \epsilon$, 
   $\alpha \rightarrow  i \alpha$, and
   $z_{j} \rightarrow \zeta_{j} = i z_{j}$), and defining the parameter
 $\displaystyle
\beta = \alpha ( 1 +  \omega^2 \epsilon^2 /2) 
$,
if follows that
 \begin{equation}
 \begin{aligned}
& K_{l+\nu}^{(N)}(r'',r';T)
=
\exp
\left[- \frac{\alpha}{2} \left( r'^{2} +  r''^{2} \right) \right]
\underbrace{\alpha^N
\,
\prod_{k=1}^{N-1}
\left[
 \int_{0}^{\infty} d r_{k} r_{k} \, e^{-\beta r_{k}^2}
\right]
\,
\prod_{j=0}^{N-1}
I_{\mu}(\alpha r_{j+1} r_{j}) 
}_{\displaystyle P_{N} (r'', r'; \alpha, \beta) }
\; .
\end{aligned}
\label{eq:propagator_RHO_orderN}
\end{equation}
In Eq.~(\ref{eq:propagator_RHO_orderN}), 
the functional form of the propagator is governed by
\begin{equation}
P_{N} (r_{N}, r_{0}; \alpha, \beta)
=
\frac{\alpha}{\gamma	_{N}}
I_{\mu}
\left( \frac{\alpha}{\gamma_{N}}   r_{N} r_{0} \right) 
\,
\exp
\left[ \frac{\alpha}{2 \lambda_{N}} \left( r_{0}^{2}+ r_{N}^{2} \right) \right]
\; ,
\label{eq:P_N-form}
\end{equation}
where the $N$-dependent parameters $\gamma_{N}$ and $\lambda_{N}$
are functions of 
$\eta=\beta/\alpha= 1 +  \omega^2 \epsilon^2 /2$;
 (i.e., they depend on the product $\omega \epsilon$).
Now, Eq.~(\ref{eq:P_N-form})
can be established by repeated application of Weber's second exponential integral for Bessel functions
(Sec.~13.31 of Ref.~\cite{wat:44} and 10.22.67 of Ref.~\cite{NIST:2010}),
 \begin{equation}
\int_{0}^{\infty} 
\exp \left( - c^2 x^2 \right)   J_{\mu} (ax) \, J_{\mu} (bx) \, x dx
=
\frac{1}{2 c^2 } 
\exp \left( - \frac{ a^2 + b^2 }{ 4c^2}  \right)  I_{\mu}  \left( \frac{ ab}{2 c^2 }   \right)
\label{eq:Weber-2nd-exponential-int}
\end{equation}
[$\Re ( \mu) > -1$ and $|\arg ( c)| < \pi/4$].
Basically, the repeated integrals in Eq.~(\ref{eq:propagator_RHO_orderN})
give an integral relation between $P_{N}$ and  $P_{N-1}$, from which 
Eq.~(\ref{eq:Weber-2nd-exponential-int}) provides three consistency conditions:
the basic recursion relation, $
\lambda_{N}
+
1/\lambda_{ N-1 } = 2 \eta$;
the relation between the parameters, 
$\lambda_{N} = \gamma_{N} /\gamma_{N-1}$;
and the additional recursion relation
$1+ 
\gamma_{N} \gamma_{N-2} =
\gamma_{N-1}^2
$.
The latter can be shown to be equivalent to the basic recursion relation,
and to the corresponding simpler relation for $\gamma_{N}$:
$
\gamma_{N} + \gamma_{N-2} =
2 \eta \gamma_{N-1}
$.
The initial values 
for the two-term and three-term relations satisfied by $\lambda_{N}$ and $\gamma_{N}$ respectively are
 $\lambda_{1} = \infty$ (or $\lambda_{2}= 2 \eta$) and
 $\gamma_{1} = 1$ with $\gamma_{2}= 2 \eta$.
The consistency of the recursion relations and generic patterns can be proved by mathematical induction, 
and a closed solution can be derived in the continuum limit $N \rightarrow \infty $.
An efficient approach~\cite{Goovaerts:1975} to derive the expressions in this limit is via a function
$\Phi (T)
= \lim_{N \rightarrow \infty}
\Phi_{N} $, 
where 
$ \Phi_{N} = \epsilon \gamma_{N}$,
with finite differences
$ \dot{\Phi}_{N} = 
(\Phi_{N+1}  - \Phi_{N} )/\epsilon
=
(\gamma_{N+1}  - \gamma_{N} )
$, 
and 
$ \ddot{\Phi}_{N} = 
(\Phi_{N+1}  +\Phi_{N-1}  - 2\Phi_{N} )/\epsilon^2
=
(\gamma_{N+1}  +\gamma_{N-1}  - 2\gamma_{N} )/\epsilon
 = \omega^2 
\Phi_{N}$.
This shows,
from Eqs.~(\ref{eq:propagator_RHO_orderN}) and (\ref{eq:P_N-form})
 that
\begin{eqnarray}
K_{l+\nu} (r'',r';T) 
=
\frac{M }{\hbar \, \Phi (T)} 
\,
\sqrt{ r' r''} 
\,
\exp \left[ - \frac{ M  }{2 \hbar } \, \frac{ \dot{\Phi} (T)}{ \Phi (T)} 
 \left(r'^{2} + r''^{2} \right) \right]
I_{\mu} \left( \frac{M }{\hbar \, \Phi (T)}  r'r'' \right)
\;   ,
\label{eq:propagator_RHO_Phi}
\end{eqnarray}
where, from the finite-difference form of the recursion relations and comparison with the free particle, the function
\begin{equation}
\Phi =  \frac{1}{\omega} \, \sinh (\omega T)
\end{equation}
(Euclidean-time version)
can be found as the solution to
the differential equation
$\ddot{\Phi} (T) = \omega^2 
\Phi (T)$ with the initial conditions
$
\Phi (0) = 0$ and
$\dot{\Phi} (0)= 1$.
In conclusion, these equations, after conversion to real time,
 show that the path integral is given by Eq.~(\ref{eq:propagator_RHO}).
 The propagator~(\ref{eq:propagator_RHO_Phi}) can also be generalized to account for possible 
 time dependence of the oscillator parameters~\cite{Goovaerts:1975}.

For the case of the inverse square potential alone, an exact derivation from a perturbative series is possible~\cite{Bhagwat:1989}, including a renormalized version for the strong-coupling 
regime~\cite{Cam_PI-singular-ISP} (similar to the case of the renormalized path integral for the two-dimensional delta interaction~\cite{Cam_PI-singular-2Ddelta}).

\section{Fourier Method for Continuous-Spectrum Operators}
\label{app:Fourier}

This appendix introduces a novel technique 
to derive the eigenfunctions of operators with a continuum spectrum. 
The general properties of this technique are discussed, and then further expanded with its relationship to Green's functions.
This topic is critical for the study of the spectral properties of the hyperbolic generators, including $S$; and it also
provides one of the various approaches to understand parabolic generators.

\subsection{Fourier Method}

For any operator $\tilde{H}$ with a purely continuous spectrum, the spectral decomposition takes the form
\begin{equation}
K_{l+\nu} (r'',r';T)
=
 \int_{\mathcal S}
  dE e^{-iET/\hbar} \, 
  \mathcal{U}_{E,l} (r'') \, \mathcal{U}^{*}_{E,l} (r')
\; ,
 \label{eq:propagator_generic_spectral-decomposition} 
  \end{equation}
where the ``energy'' values $E \equiv \tilde{E}$ of $\tilde{H}$ extend over the set ${\mathcal S}$.
We assume that $\tilde{H}$ has a purely continuous spectrum, and can be regarded
as a sort of Hamiltonian operator associated with an effective time evolution, i.e., defining 
 $ K_{(d)} ({\bf r''},{\bf r'};T) =  \bra{\bf r''}e^{-i\tilde{H}T}\ket{\bf r'}$ and
 extracting the radial counterpart 
 according to the rules of Sec.~\ref{sec:PI-setup_CQM}.
 
Equation~(\ref{eq:propagator_generic_spectral-decomposition}) has the form 
of a Fourier transform restricted to the set ${\mathcal S}$.
Its inverse Fourier transform 
can be obtained by performing the integral with respect to the variable $T \in (-\infty, \infty)$,
with the kernel
$ \exp \left( {iET}/{\hbar} \right)$. This general inverse Fourier integral converts the propagator $ K_{l+\nu} (r'',r';T)$ into
the wave-function product 
$  \mathcal{U}_{E,l} (r'') \, \mathcal{U}^{*}_{E,l} (r')$, i.e.,
\begin{equation}
F(E; r'',r') \equiv   \mathcal{U}_{E,l} (r'') \, \mathcal{U}^{*}_{E,l} (r')
=
\frac{1}{2 \pi \hbar} \, \int_{-\infty}^{\infty} 
dT \, \exp \left( \frac{iET}{\hbar} \right) \, K_{l+\nu} (r'',r';T)
\; ,
\label{eq:eigenfunctions_from-Fourier}
\end{equation}
where the values of the physical energy on the real axis are restricted to the original set $E \in {\mathcal S}$.
Indeed, this general theorem
simply follows by performing the inverse Fourier integral on the right-hand side of
Eq.~(\ref{eq:propagator_generic_spectral-decomposition}) written in terms of a variable
 $E'$, with the familiar auxiliary identity
 $\int_{-\infty}^{\infty} d T e^{i(E-E') T/\hbar} = 2 \pi \hbar \delta (E-E')$.
 
In conclusion,
Eqs.~(\ref{eq:propagator_generic_spectral-decomposition})
and (\ref{eq:eigenfunctions_from-Fourier})
provide the broader context and justification for the results of 
Subsec.~\ref{subsec:Fourier-method} and 
Sec.~\ref{sec:spectral_CQM_hyperbolic}.
This ``Fourier method'' is remarkably simple and elegant, but it has not been explicitly used in the literature.
This is partly due to the fact that operators with a purely continuous spectrum are not common
 (other than the trivial case of a free particle),
and partly because it is customary to use Green's functions techniques, which are related but not identical
to the result of Eq.~(\ref{eq:eigenfunctions_from-Fourier}).

\subsection{Green's Functions: Definitions and Relation to the Fourier Method}

 In general,
the retarded/advanced Green's functions or resolvents associated with $\hat{H}$ are defined 
from
\begin{equation}
G^{(\pm)}_{l+\nu}(r'',r'; E)
=
\pm
\frac{1}{i \hbar} \, \int_{-\infty}^{\infty} 
dT \, \theta (\pm T) \, \exp \left( \frac{iET}{\hbar} \right) \, K_{l+\nu} (r'',r';T)
\; ,
\label{eq:radial-Green-function}
\end{equation}
where $\theta $ stands for the Heaviside function, and with the replacement 
 $E \rightarrow E \pm i 0^{+} $ that guarantees convergence.
Unlike Eq.~(\ref{eq:eigenfunctions_from-Fourier}), the Green's function technique does not give 
a direct result for the wave function product, though this product can be extracted via an additional step
as the residue of the energy poles.

The definitions of Eq.~(\ref{eq:radial-Green-function}) 
correspond to the Fourier transform of the (retarded/advanced) Green operators
\begin{equation}
G^{(\pm)} (T) =  \theta (\pm T) \, e^{-i \hat{H} T/\hbar}
\; .
\label{eq:Green-operators}
\end{equation}
This should be contrasted with the propagator counterparts that involve 
 the time evolution operator
$U = e^{-i \hat{H} T/\hbar}$
of Eq.~(\ref{eq:propagator_QM_from_evolution_operator}) without the Heaviside cutoff. 
Equation~(\ref{eq:Green-operators})
yields the Fourier-transformed or energy Green operators
\begin{equation}
G^{(\pm)} (E)= \left( E - \hat{H} \pm i 0^{+} \right)^{-1}
\; ,
\label{eq:Green-operators-energy}
\end{equation}
where the $i 0^{+}$ prescription provides convergence for each case. 
The corresponding coordinate-space representations give the associated time and energy Green's functions
$G^{(\pm)}_{(d)} (r'',r';T) $
and
$G^{(\pm)}_{(d)} (r'',r';E) $; in particular,
\begin{equation}
G^{(\pm)}_{(d)} ({\bf r''},{\bf r'};E) =
\left\langle {\bf r''}
\left|
\left( 
E - \hat{H} \pm i 0^{+} 
\right)^{-1}
 \right|
{\bf r'}
\right\rangle   
\label{eq:GF_as_operator}
\; .
\end{equation}
Equations~(\ref{eq:Green-operators})--(\ref{eq:Green-operators-energy})
can be applied either to the full-fledged multidimensional quantities, or their reduced radial counterparts,
which follow from the usual hyperspherical expansion; specifically,
the radial energy Green's functions $G_{l +\nu}(r'',r';E)$ in Eq.~(\ref{eq:radial-Green-function})
are  defined from 
\begin{equation}
G^{(\pm)}_{(d)}({\bf r''}, {\bf r'}; E) 
=
\left( r'' r' \right)^{-(d-1)/2}
\,
\sum_{l= 0}^{\infty}
\sum_{m=1}^{d_{l}}
Y_{l m} ({\bf \Omega''})
Y_{l m}^{ *} ({\bf \Omega'})
G^{(\pm)}_{l +\nu}(r'',r';E)
\;  .
\label{eq:Green-function_partial_wave_exp}
\end{equation}

Moreover, with the usual distributional expansion 
$\left( A \pm i 0^{+} \right) ^{-1} = \mathcal{P}(A^{-1}) \mp i \pi \delta (A)$ (where 
$\mathcal{P}$ is the Cauchy principal value), the ratio
$-  \left( G^{(+)}(E) -  G^{(-)}(E) \right)/(2 \pi i) = \delta (E- \hat{H} )$ is a state density operator.
Thus, for a continuous-energy spectral expansion,
\begin{equation}
\begin{aligned} 
F(r'',r'; E) \equiv 
  \mathcal{U}_{E,l} (r'') \, \mathcal{U}^{*}_{E,l} (r')
& = 
\frac{1}{2 \pi \hbar} \, \int_{-\infty}^{\infty} 
dT \, \exp \left( \frac{iET}{\hbar} \right) \, K_{l+\nu} (r'',r';T)
\\
& 
=
- \frac{1}{2 \pi i} \, 
\underbrace{ 
\left[  G^{(+)}_{l+\nu} (r'',r'; E) - G^{(-)}_{l+ \nu}(r'',r'; E) \right] }_{ \displaystyle \text{disc} [G_{l+\nu} (r'',r';E)] 
}
\; ,
\label{eq:eigenfunctions_from-Green}
\end{aligned}
\end{equation}
where $\text{disc} [ G_{l+\nu} (E) ] = G^{(+)}_{l+\nu}(E) - G^{(-)}_{l+\nu}(E) $ 
measures the discontinuity of the Green's functions
across the branch cut in the complex energy plane.
In short,
Eq.~(\ref{eq:eigenfunctions_from-Green}) summarizes the connection between the straightforward 
Fourier transform of the propagator and the Green's functions; for the former, the time domain of the
Fourier integrals involves the whole time axis, while for the Green's functions, the time domain is restricted 
to the positive and negative half-axes.

\subsection{Green's Functions---Example: Parabolic Generators}

As an example of the relations above, we can revisit the conformal operator $H$ and its parabolic class.
The whole time-axis Fourier integral gives directly the wave function 
product~(\ref{eq:eigenfunctions_from-Fourier_ISP_explicit}), which we derived from
Eq.~(\ref{eq:Bessel-product-integral}).
If we instead use the related integrals of 
Eq.~(\ref{eq:Bessel-Hankel-product-integrals}),
the Green's functions become 
\begin{equation}
G^{(\pm)}_{l+\nu} (r'',r'; E) 
= \mp \pi i \,  (M/\hbar^2) 
\, \sqrt{r' r''} 
J_{\mu} (kr_{<}) H^{(1,2)}_{\mu} (kr_{>}) 
\; ,
\label{eq:Green-functions_H-op}
\end{equation}
where $r_{<}$ ($r_{>}$) is the lesser (greater) of $r'$ and $r''$; then,
\begin{displaymath}
G^{(+)}_{l+\nu}  (r'',r'; E)  - G^{(-)}_{l+\nu}  (r'',r'; E)  = - \pi i \,  
 \frac{M}{\hbar^2} 
\, \sqrt{r' r''} 
J_{\mu} (kr_{<}) \left[ H^{(1)}_{\mu} (kr_{>}) + H^{(2)}_{\mu} (kr_{>}) \right] 
\; ,
\end{displaymath}
which, with $ H^{(1)}_{\mu} (z) + H^{(2)}_{\mu} (z) = 2 J_{\mu} (z)  $,
gives $- 2 \pi i $
times the wave function product displayed in Eq.~(\ref{eq:eigenfunctions_from-Fourier_ISP_explicit}),
in agreement with the relation~(\ref{eq:eigenfunctions_from-Green}).

In the next section, we will apply the technique of Eq.~(\ref{eq:eigenfunctions_from-Fourier})
to derive the spectral decomposition associated with the more involved
 conformal operator $S$, and will display the corresponding network of relations developed here.

\section{Relevant Integral Representations for the Product of Bessel and Whittaker Functions as Fourier Transforms 
of Propagators}
\label{app:Bessel-Whittaker-products}

In this appendix, we summarize the key results on integral representations of products 
of the special functions relevant for the propagators discussed in this paper. 
We will first identify the identities relevant for the operator $H$ and related parabolic generators, i.e., 
generally for the pure inverse square potential, using products of Bessel functions; 
and we will then explore a general method and appropriate identities 
for the hyperbolic operators, corresponding to an inverted radial harmonic oscillator, using products of Whittaker functions.
The identification of the relevant identities is a useful addition to the literature of path integrals; and, in the case of the latter class, we establish, inter alia, a new integral identity.

\subsection{Integral Representations for the Product of Bessel Functions}

Let us consider the following integral representation of the product of Bessel functions (Sec.~13.7
 of Ref.~\cite{wat:44}):
\begin{equation}
J_{\mu} (z')  J_{\mu} (z'') 
= \frac{1}{2\pi  i}
\int_{c-i\infty}^{c+ i \infty}
\exp \left( \frac{s}{2 } - \frac{z'^{\, 2} + z''^{\, 2} }{2s} \right) 
\, I_{\mu} \left( \frac{z' z'' }{s} \right) \frac{ds}{s}
\; ,
\label{eq:Bessel-product-integral}
\end{equation}
where $c$ is a real positive constant and $\Re (\mu) > -1$.
As shown in Ref.~\cite{wat:44}), this representation~(\ref{eq:Bessel-product-integral})
can be established by combining an appropriate form of Bessel function Gegenbauer addition theorems 
with Schl\"{a}fli's integral 
$\displaystyle J_{\nu}\left(z\right)=\frac{(\tfrac{1}{2}z)^{\nu}}{2\pi i}\int_{-\infty}^{(0+)
}\exp\left(t-\frac{z^{2}}{4t}\right)\frac{\mathrm{d}t}{t^{\nu+1}}$.
As shown in Subsec.~\ref{subsec:spectral_H}, we consider the limit 
 $c \rightarrow 0^{+}$, that reduces the integral~(\ref{eq:Bessel-product-integral}), with $s=c + 2 i t$, 
 to the form~(\ref{eq:Bessel-product-integral-2}),
where $ t \rightarrow t- i 0^{+} $.
Its relevance for our study stems from the fact that the substitution
$t \equiv ET/\hbar$
makes it directly applicable to Eq.~(\ref{eq:eigenfunctions_from-Fourier_ISP}) for the operator $H$ 
(inverse square potential) as a Fourier transform,
leading to Eq.~(\ref{eq:eigenfunctions_from-Fourier_ISP_explicit}).

 In a similar way, the counterparts of Eq.~(\ref{eq:Bessel-product-integral}) for the half-axis intervals can be
 found using Hankel functions (for which the corresponding Schl\"{a}fli's integrals only involve half-axes),
 yielding
 \begin{equation}
H^{(1,2)}_{\mu} (z_{>}) \, J_{\mu} (z_{<}) 
= \pm \frac{1}{\pi  i}
\int_{0}^{c \pm  i \infty}
\exp \left[ \frac{s}{2 } - \frac{(z'^{\, 2} + z''^{\, 2}) }{2s} \right]
\, I_{\mu} \left( \frac{z' z'' }{s} \right) \frac{ds}{s}
\; 
\label{eq:Bessel-Hankel-product-integrals}
\end{equation}
 where $c$ is again a real positive constant and $\Re (\mu) > -1$; and
  $z_{>}$  and $z_{<}$
are the greater and lesser of the set $\{z', z''\}$ respectively.
(A related integral is listed in Ref.~\cite{Gradshteyn-Ryzhik:2000}, 6.653-1.)

\subsection{Integral Representations for the Product of Whittaker Functions}

We will begin by summarizing and adapting the technique defined in 
Ref.~\cite{Buchholz:1969} (Section 6) to set up integral representations of products of Whittaker functions.
The method involves writing a single Whittaker function as a confluent hypergeometric function that is 
related to a Bessel function in integral form. Writing this integral twice for a product of two Whittaker functions gives
 \begin{equation}
  \begin{aligned}
\mathcal{M}_{\lambda_{1},\mu_{1}/2}(z_{1})     \,
\mathcal{M}_{\lambda_{2},\mu_{2}/2}(z_{2})
& =
 \frac{ 4 \, \left( z_{1} z_{2} \right)^{1/2}
 e^{ ( z_{1} + z_{2} )/2}  }{ \Gamma \left( \frac{1+\mu_{1}}{2} + \lambda_{1} \right)
 \, \Gamma \left( \frac{1+\mu_{2}}{2} - \lambda_{2} \right) }
\\
& \qquad 
 \times \int_0^{\infty} dt \int_0^{\infty} du \,
e^{-t^2-u^2} 
 t^{2 \lambda_{1}} u^{2 \lambda_{2}}
J_{\mu} ( 2t\sqrt{z_{1}} )
J_{\mu}( 2u\sqrt{z_{2}} )
\; ,
  \end{aligned}
  \label{eq:Whittaker-product-int_1}
      \end{equation}
whence a large class of representations can be developed by appropriate variable substitutions.
For the important case of products of functions with the same indices, choosing 
$\lambda_{1} =- \lambda_{2} \equiv \lambda$ and $\mu_{1}= \mu_{2} \equiv \mu $,
the following representation is obtained by going to ``polar coordinates'' via
$t= \rho \cos \phi$ and $u= \rho \sin \phi$, and evaluating the integral with respect to $\rho$ 
 via Weber's second exponential integral~(\ref{eq:Weber-2nd-exponential-int}), 
 \begin{equation}
  \begin{aligned}
 \! \! \!
 \mathcal{M}_{i \kappa,\mu/2} (- ix' )   \, &
 \mathcal{M}_{ -i \kappa,\mu/2}  ( ix'' )
 =
 \frac{ 2 \, \left( x' x'' \right)^{1/2}
 e^{ i ( x' - x'' )/2}  }{ \Gamma \left( \frac{1+\mu }{2} +i \kappa\right)
 \, \Gamma \left( \frac{1+\mu }{2} - i \kappa \right) }
  \\
& \qquad 
\times \int_0^{\frac{\pi}{2}}
 d\phi \,
 \left( \cot{\phi} \right)^{2i\kappa}
e^{i ( x' \cos^2{ \phi}- x'' \sin^2{ \phi} ) }
 \,
 I_{\mu} \left( \sqrt{ x' x'' }  \, \sin{2\phi}  \right)
 \; ,
  \end{aligned}
    \label{eq:Whittaker-product-int_2}
    \end{equation}
   where we are also making the assignments 
  $\lambda = i \kappa$, $z_{1} = -ix'$, and $z_{2} = ix''$,
   tailored specifically to the Whittaker functions with imaginary first index needed 
   for the radial inverted oscillator or the conformal operator $S$.
   
  From Eq.~(\ref{eq:Whittaker-product-int_2}), a whole class of integrals can be obtained by appropriate substitutions.
  In particular, if $\zeta$ is defined such that $\sin 2 \phi = 1/(i \sinh \zeta)$, then 
Eq.~(\ref{eq:Whittaker-product-int_2}) turns into a form that appears to match the propagator of Eq.~(\ref{eq:propagator_inverted-RHO}), with $\zeta = \omega t$.
However, this is a nontrivial substitution that takes the variable $\zeta $ into the complex plane and requires further analysis. We will first implement this substitution in two steps: (i) defining a real variable $s$ such that 
$\sin 2 \phi = 1/\cosh s$; (ii) replacing $s$ by $\zeta$ via $\cosh s = i \sinh \zeta$ in the complex plane. Moreover, there are issues with the interpretation of the integral in the complex plane that require an
appropriate deformation of the integration contour. This is shown next.

First, from the range of the integral~(\ref{eq:Whittaker-product-int_2}), the substitution 
 $\sin 2 \phi = 1/\cosh s$ establishes a one-to-one correspondence that
maps $2 \phi \in [0, \pi]$ into $s \in (-\infty, \infty)$, only involving real variables.
By straightforward algebra, 
$d \phi = ds/(2 \cosh s)$, 
 $\cos2 \phi = \tanh s$,  and $\cot \phi = e^{s}$, 
 which directly yields an integral of the well-known form
 of Eq.~(3a) in Sec.~6.1 of Ref.~\cite{Buchholz:1969} in the specific variant
  \begin{equation}
  \begin{aligned}
 \! \! \!
 \mathcal{M}_{i \kappa,\mu/2} (- ix' )   \, &
 \mathcal{M}_{- i \kappa,\mu/2}  ( i x'' )
 =
 \frac{  \left( x' x'' \right)^{1/2}
   }{ \Gamma \left( \frac{1+\mu }{2} +i \kappa\right)
 \, \Gamma \left( \frac{1+\mu }{2} - i \kappa \right) }
  \\
& \qquad 
\times \int_{-\infty}^{\infty}
 \frac{d s}{\cosh s}  \,
\exp \left( 2 i \kappa s \right)
\exp \left[ \frac{i}{2} \left( x' + x'' \right) \tanh s \right]
 \,
  I_{\mu} \left( \frac{\sqrt{ x' x'' } }{\cosh s}  \right)
 \; 
  \end{aligned}
    \label{eq:Whittaker-product-int_3}
    \end{equation}
 (which agrees with Ref.~\cite{Gradshteyn-Ryzhik:2000}, 6.669-5).

 Second, the transformation 
$\cosh s = i \sinh \zeta$
is evidently a translation $\zeta = s + \sigma$ along the imaginary axis in the complex plane,
with displacement $\sigma$.
The general solution of this equation
is  $\zeta = \pm s - i \pi/2 + 2 \pi n$, with $n \in \mathbb{Z}$,  which follows from the $2 \pi i$ 
periodicity of the $\cosh$ function (or simply by finding the solution via inversion of
the hyperbolic or exponential functions). Moreover,
the negative sign can be ignored if we only consider the case of a path without inversion.
This shows that the translation displacement is $\sigma =\zeta_{n} \equiv \zeta_{0} + 2\pi n$, 
with the ``principal value'' $\zeta =i c$, with $c= - \pi/2$, which we will use below.
With this transformation, the following relations are immediately satisfied:
$ds = d \zeta$,
$\sinh s = i \cosh \zeta$, 
$\tanh s =  \coth \zeta$, 
$\sinh s = i \cosh \zeta$, 
and $e^{s} = e^{i \pi/2}  e^{\zeta}$. As a result,
the integration path for the transformed version of the integral~(\ref{eq:Whittaker-product-int_3})
is an infinite straight line $L$
parallel to the real $\zeta$ axis, with  $\Im \zeta = \Im \sigma = c$.
With this procedure, we have established a novel identity 
   \begin{equation}
  \begin{aligned}
 & \! \! \!   \! \! \!   \! \! \! 
 \mathcal{M}_{i \kappa,\mu/2} ( -ix' )   \, 
 \mathcal{M}_{- i \kappa,\mu/2}  ( i x'' )
 =
 e ^{- \pi \kappa} \,
 \frac{  \left( x' x'' \right)^{1/2}
   }{ \Gamma \left( \frac{1+\mu }{2} +i \kappa\right)
 \, \Gamma \left( \frac{1+\mu }{2} - i \kappa \right) }
  \\
& \qquad   \qquad  \; \;
\times \int_{i c-\infty}^{i c + \infty}
 \frac{d  \zeta}{i \sinh \zeta}  \,
\exp \left( 2 i \kappa \zeta \right)
\exp \left[ \frac{i}{2} \left( x' + x'' \right) \coth \zeta \right]
 \,
  I_{\mu} \left( \frac{ \sqrt{ x' x'' } }{i\sinh \zeta}  \right)
 \; .
  \end{aligned}
    \label{eq:Whittaker-product-int_4}
    \end{equation}
  The integral of direct applicability
 for the required form of the propagator~(\ref{eq:propagator_inverted-RHO}) involves using a path with 
 $c \rightarrow 0$.
  This can be justified from the following properties:
  (i) the only singularities of the integrand occur
 at $\zeta = 2 \pi n i$, which we will circumvent with appropriate integration contours; 
 and (ii) the behavior is regular at infinity along curves asymptotically parallel to the real axis, 
 with the integral vanishing exponentially for $|\Re \zeta | \rightarrow \infty$. 
 Thus,  $\oint_{C} F(\zeta) d \zeta =0$, for the integrand $F(\zeta)$ in Eq.~(\ref{eq:Whittaker-product-int_4}) 
 with a closed contour $C$ consisting of the line $L$ and a parallel line $L'$ with $c \in (-\pi/2, 0)$.
 This proves Eq.~(\ref{eq:Whittaker-product-int_4}) for arbitrary $c$ with $-\pi/2 < c < 0$.
  Moreover, if the restriction to straight lines is relaxed,
  the line $L'$ can further deformed into any other path with asymptotic limits 
  $ic'  - i \infty$ and 
  $ic'' +  i \infty$, where $c'$ and $c''$ are arbitrary.
  For the propagator~(\ref{eq:propagator_inverted-RHO}), it suffices to take the limit $c \rightarrow 0$.
   It should be noted that this is an analog for Whittaker functions of the 
 Bessel-function identity~(\ref{eq:Bessel-product-integral}).
 With this auxiliary integral~(\ref{eq:Whittaker-product-int_4}),
 making the replacements 
  $\zeta = \omega T$,  $x' = M  \omega  {r'}^{2}/\hbar $, 
 $x''= M  \omega  {r''}^{2}/\hbar $, $\kappa = E/(2 \hbar \omega)$, and  $c=0^{\mp}$, 
 the main result of Eqs.~(\ref{eq:eigenfunctions_from-Fourier_inverted-RHO_explicit-1})--(\ref{eq:eigenfunctions_from-Fourier_inverted-RHO_explicit-2}) in Sec.~\ref{sec:spectral_CQM_hyperbolic} is straightforwardly derived.
 
 An alternative set of representations can be established by performing integrals of the 
 forms~(\ref{eq:Whittaker-product-int_3}) and (\ref{eq:Whittaker-product-int_4}), but restricted
 to the half-axis intervals.
 For example,
  Eq.~(5b) in Sec.~6.1 of Ref.~\cite{Buchholz:1969} reads 
   \begin{equation}
  \begin{aligned}
\! \! \! \! \! \!   W_{\kappa,\mu/2} (a_{1} t)     \,  & \mathcal{M}_{\kappa,\mu/2}   (a_{2} t)
=
   \frac{ t \sqrt{ a_{1} a_{2} }}{ \Gamma \left( (1+\mu)/2 - \kappa \right) }
 \\
  & \; \; \;  \cdot \,  \int_{0 }^{\infty} d \xi  \, \exp \left[ \displaystyle - \frac{1}{2} \left( a_{1}+a_{2} \right) t \cosh \xi \right]
\,
I_{\mu}  \left( t \sqrt{a_{1} a_{2} }  \, \sinh \xi \right)  
\, \left[ \coth \left( \frac{\xi}{2} \right) \right]^{2 \kappa}
\; 
\label{eq:Whittaker-integral_half-line_BU}
\end{aligned}
  \end{equation}
   (which agrees with Ref.~\cite{Gradshteyn-Ryzhik:2000}, 6.669-4),
 where 
 $a_{1,2}$ are real parameters satisfying the critical inequality $a_{1}> a_{2}$, and
 $\Re ((1+\mu)/2 - \kappa) > 0$.
 The variables $t$, $a_{1}$, and $a_{2}$  in Eq.~(\ref{eq:Whittaker-integral_half-line_BU}), 
provide some flexibility of choices;
however, as shown in the steps leading to their derivation,
with the notation used in Eq.~(\ref{eq:Whittaker-product-int_1}), one can identify
$z_{1}= t a_{1}$ and $z_{2}= t a_{2}$.
Then, as before, one can make the additional replacements $\kappa \rightarrow i \kappa $ and
$z_{1,2} =- i x', - i x''$. (This could be done most easily with $t=-i$ and $a_{1,2}$ chosen from the set $ x',x''$.)
Finally, with the substitution
$\sinh \xi =1/\sinh \zeta$
(which implies 
$\cosh \xi = \coth \zeta$, $\coth (\xi/2) = e^{\zeta}$, and 
$d\xi= -d \zeta/\sinh \zeta$),
Eq.~(\ref{eq:Whittaker-integral_half-line_BU}) turns into 
  \begin{equation}
  \begin{aligned}
\! \! \! \! \! \!   W_{i \kappa,\mu/2} (i x_{>})     \,  & \mathcal{M}_{i \kappa,\mu/2} (i x_{<} )  
=
   \frac{  \sqrt{ x' x''} }{ \Gamma \left( (1+\mu)/2 - i \kappa \right) }
 \\
 & \qquad
\times \int_{0}^{\infty}
 \frac{d  \zeta}{i \sinh \zeta}  \,
\exp \left( 2 i \kappa \zeta \right)
\exp \left[ \frac{i}{2} \left( x' + x'' \right) \coth \zeta \right]
 \,
  I_{\mu} \left( \frac{ \sqrt{ x' x'' } }{i\sinh \zeta}  \right)
\; ,
\label{eq:Whittaker-integral_half-line_main}
\end{aligned}
  \end{equation}
where 
 $x_{>}$  and $x_{<}$
are the greater and lesser of the set $\{x', x''\}$ respectively.

Incidentally, the known identity of Eq.~(\ref{eq:Whittaker-integral_half-line_BU}) is proved, as stated in Ref.~\cite{Buchholz:1969}, by using appropriate substitutions combined with identities
(semi-circuital and connection) among Whittaker functions.
In fact, this is a reversal of the procedure we used in the main text; in other words, we could just as well use the 
symmetry property~(\ref{eq:I_pm-symmetries}), 
along with the Whittaker identities~(\ref{eq:Whittaker-M-analytic-cont}) and (\ref{eq:Whittaker-M-W-connection}),
and with appropriate limits, to justify the half-axis integrals(s)~(\ref{eq:Whittaker-integral_half-line_main})
 from the full integral~(\ref{eq:Whittaker-product-int_4});
and from Eq.~(\ref{eq:Whittaker-integral_half-line_main}) rederive 
Eq.~(\ref{eq:Whittaker-integral_half-line_BU}) via $\sinh \xi =1/\sinh \zeta$.

\section{Generalized Symmetry Generators of CQM---Differential Equation
and Consistency Checks}
\label{app:consistency-check_diff-eq-S}

In this appendix, we consider another aspect of the CQM generalized generator $G= uH + vD + w K$:
 its behavior and spectral properties using a differential equation approach. 
These properties can be deduced by going to the Schr\"{o}dinger picture for 
the general multicomponent case of the dAFF model we introduced in Sec.~\ref{sec:symmetries}
($d$ components as $d$-dimensional position coordinates in quantum mechanics). 
This setup for $d$ dimensions and for the generic operator $G$ extends the particular results
of Ref.~\cite{AFF:76}.

Then, in the Schr\"{o}dinger picture, with the Hamiltonian 
$\tilde{H}_{G} (\mathbf{r}, \mathbf{p} ) \equiv G/\sigma$
introduced in Eq.~(\ref{eq:G-as-Hamiltonian_reduced}),
the states $ \Psi ({\mathbf r}, \tau) $ evolve according to
\begin{equation}
i \hbar \, \frac{ \partial \Psi ({\mathbf r}, \tau)}{\partial \tau}
=
\tilde{H}_{G} \left( {\mathbf r}, {\mathbf p}  \right) \, 
\Psi ({\mathbf r}, \tau)
\; .
\label{eq:Schrodinger-eq}
\end{equation}
From Eq.~(\ref{eq:Schrodinger-eq}),
 ``stationary states'' can be defined with respect to $\tau$, such that the eigeinvalue equation 
$\tilde{H}_{G} \ket{\psi} = \tilde{E}_{G} \ket{\psi}$ is satisfied; this 
leads to the radial-coordinate differential equation
\begin{equation} 
\frac{1}{2} \, \frac{\hbar^{2}}{M} \, 
\left [ -\frac{d^2}{dr^2}
+ \frac{g +(l+\nu)^{2}  - 1/4 }{ r^2 } + 
  \frac{M^{2} }{\hbar^{2}} 
\left(-\frac{\Delta}{4} \right)  r^2 \right]
\mathcal{U}_{ \mathfrak{g},l } (r)
= 
\tilde{E}_{G}
\, 
\mathcal{U}_{ \mathfrak{g},l } (r)
\; ,
\label{eq:gen-generator_diff-eq}
\end{equation}
with the eigenvalues of the operator $G$ being 
$\hbar \mathfrak{g} = \sigma \tilde{E}_{G}$---see Eq.~(\ref{eq:eigen-g_E}) and the discussion therein.
In Eq.~(\ref{eq:gen-generator_diff-eq}), the reduced function $\mathcal{U}_{ \mathfrak{g},l } (y)$
corresponds to the
$d$-dimensional wave function 
$\psi ({\mathbf r}) = r^{-(d-1)/2} \,  \mathcal{U}_{ \mathfrak{g},l } (r)
\, Y_{l \boldsymbol{m}} ({\bf \Omega})$,
following the usual separation of angular variables.

With the substitutions $A= - i (M/\hbar )\sqrt{ \Delta }/2$,
$\lambda =  i \sigma \mathfrak{g}/\sqrt{\Delta} $,
 and  $\mu = g +(l+\nu)^{2}$,
Eq.~(\ref{eq:gen-generator_diff-eq}) is of the known form (Ref.~\cite{Buchholz:1969}, p. 34)
\begin{equation} 
\left[-\frac{d^2}{dr^2} 
+\frac{\mu^2-1/4}{r^2}
+A^2 r^2\right] \Phi (r)
=4 \lambda A \, \Phi (r) 
\; ,
\label{eq:Whittaker-transformed-DE-HO}
\end{equation}
which can be solved exactly in terms of Whittaker functions. Specifically,
 the standard Whittaker differential equation for a function $P (z)$ can be transformed into 
Eq.~(\ref{eq:Whittaker-transformed-DE-HO}) with the substitution $z = A r^{2} $,
such that the solutions are 
\begin{equation} 
\Phi (r)=r^{-1/2} \, P_{\lambda,\mu/2}(Ar^2) 
\; ,
\label{eq:Whittaker-transformed-DE-HO_sol}
\end{equation}
where, with the notation of Ref.~\cite{Buchholz:1969},
$P_{\lambda,\mu/2}(z)$ stands for 
either one of the Whittaker functions
$\mathcal{M}_{\lambda,\mu/2}(z) $
and
 $W_{\lambda,\mu/2}(z) $,
or a linear combination thereof. 
The substitutions used in Eq.~(\ref{eq:Whittaker-transformed-DE-HO}), which involve the scale $A$ and the index 
$\lambda$, are completely general, covering all classes of operators (elliptic, parabolic, and hyperbolic), according to the
value of $\Delta$.

Furthermore,
enforcing the regular behavior at the origin, the function $W_{\lambda,\mu/2}(z) $ should be excluded 
for the wave function solutions---though it is still relevant for the Green's functions (see below).
Then, using the values for $A$, $\lambda$, and $\mu$, 
the ensuing regular choice of the solution~(\ref{eq:Whittaker-transformed-DE-HO_sol})
to Eq.~(\ref{eq:gen-generator_diff-eq}) becomes
\begin{equation} 
\mathcal{U} (r) 
=  r^{-1/2} \,
 \mathcal{M}_{ i  \sigma \mathfrak{g} /\sqrt{\Delta},\mu/2} 
 ( - i  \check{r}^{2} ) 
\; ,
\label{eq:Whittaker-transformed-DE-HO_sol_2}
\end{equation}
where
\begin{equation} 
  \check{r}^{2} = \frac{\sqrt{\Delta}}{2} \, \left(\frac{M}{\hbar} \right) r^{2}
  \equiv \frac{M |\omega|}{\hbar}  r^{2}
\; .
\label{eq:dimensionless-r}
\end{equation}
The variable $ \check{r}$ in Eq.~(\ref{eq:dimensionless-r})
is written in terms of the general CQM frequency of
Eq.~(\ref{eq:squared-frequency}), i.e., $\omega = - i |\omega|$, which is in agreement 
with the definitions in 
Sec.~\ref{sec:spectral_CQM_hyperbolic}:
in Eq.~(\ref{eq:dimensionless-parameters-I}) and with the 
analytic extension $\omega \rightarrow - i \omega= - i\sqrt{\Delta}/{2}$ used therein.
Moreover, the solution~(\ref{eq:Whittaker-transformed-DE-HO_sol_2}) gives a continuous spectrum
with $ \mathfrak{g} \in  (-\infty, \infty )$.
This set of results reproduces the path-integral treatment leading to Eq.~(\ref{eq:inverted-RHO_energy-eigenfunctions-1})
[with Eq.~(\ref{eq:dimensionless-parameters-I})]
for the hyperbolic-operator eigenfunctions.
Most importantly, Eq.~(\ref{eq:Whittaker-transformed-DE-HO_sol_2}) is completely general; when $\Delta <0$, it
can also be applied to the elliptic case---this amounts to the analytic
continuation $\kappa =  i  \sigma \mathfrak{g}/\sqrt{\Delta} \rightarrow  \sigma \mathfrak{g}/\sqrt{\Delta} $,
which (from the asymptotics)
requires eigenvalue ``quantization'' leading to the generalized Laguerre polynomials,
 as in Eq.~(\ref{eq:tilde-H_eigenstates}). For the parabolic case, the limit $\Delta =0$ leads to Bessel functions 
 $J_{\mu} (kr)$ as regular solutions, as can also be verified directly from Eq.~(\ref{eq:gen-generator_diff-eq}).
 In the latter case, there is a known subtlety or ambiguity in the choice of the sign of the regular solution 
 from the set $J_{\pm \mu}(kr)$, with a physical argument~\cite{Landau:77} selecting the positive sign---this 
 is related to the inequivalence of the self-adjoint extension method with physical regularization 
 techniques~\cite{cam_inequivalence}.
 In short, the differential-equation approach fully agrees with the path-integral results of the main text
for all the conformal generators.

In addition, the Green's functions can be derived from the corresponding 
differential equation~(\ref{eq:gen-generator_diff-eq}), 
using the general solutions we already found, Eq.~(\ref{eq:Whittaker-transformed-DE-HO_sol}). 
The usual notational simplification $E \equiv \tilde{E}_{G}$ is used below
[or else, this symbol could be replaced in favor of $\mathfrak{g}$, according to Eq.~(\ref{eq:eigen-g_E})].
The radial energy Green's functions $G^{(\pm)}_{l +\nu}(r'',r';E)$ are defined in
Eq.~(\ref{eq:propagator_partial_wave_exp}),
which we will rescale as ${\mathcal G}^{(\pm)}_{l +\nu}(r'',r';E)= (\hbar^{2}/2M) \, G^{(\pm)}_{l +\nu}(r'',r';E)$.
As is well-known, because of the specific form of 
Eqs.~(\ref{eq:Green-operators-energy})--(\ref{eq:GF_as_operator}), which effectively 
invert the Schr\"{o}dinger eigenvalue equation,
these coincide with the Green's functions used for the solution of linear differential equations.
Then, the radial differential equation for ${\mathcal G}_{l +\nu}(r'',r';E)$ reads
\begin{equation}
\left\{ \frac{d^{2}}{dr'^{2}} + 
\frac{2M}{ \hbar^{2} }
\left[
E - V( r')
\right]
-
\frac{ \left( l + \nu \right)^{2}  
- 1/4}{r'^{2}}
\right\}  
{\mathcal G}_{l+\nu}(r'',r';E)
=
\delta ( r''- r')
\;  ,
\label{eq:GF_radial}
\end{equation}
which is a particular case of a one-dimensional Sturm-Liouville problem (with constant coefficient $p(r) =1$
for the second-order derivative)~\cite{Stakgold}. 
Equation~(\ref{eq:gen-generator_diff-eq})
is the homogeneous form of Eq.~(\ref{eq:GF_radial}), with the obvious identifications;
then, defining the the functions
$\mathcal{U}^{(\pm)}_{(<)}(r)$
and 
$\mathcal{U}^{(\pm)}_{(>)} (r)$ 
that satisfy boundary conditions at the left boundary (here: $r=0$) and right boundary (here: $r=\infty$),
 then~\cite{Stakgold}
\begin{equation}
{\mathcal G}^{(\pm)}_{l +\nu}(r'',r';E)
=
\frac{ \mathcal{U}^{(\pm)} _{(<)}  (r_{<}) \;
\mathcal{U}^{(\pm)} _{(>)} (r_{>}) 
}{ 
p (r') \;  
\mathfrak{W} \left\{ 
\mathcal{U}^{(\pm)} _{(<)}  \, , \,
\mathcal{U}^{(\pm)} _{(>)} 
\right\} 
(r') }
\;  ,
\label{eq:GF_radial_explicit}
\end{equation}
where 
 $r_{<}$ ($r_{>}$) is the lesser
(greater) of $r'$ and $r''$ and
 $\mathfrak{W} \left\{ \mathcal{U}^{(\pm)}_{(<)} ,\mathcal{U}^{(\pm)}_{(>)}  \right\}$
is the Wronskian of 
$ \mathcal{U}^{(\pm)}_{(<)} (r)$
and 
$ \mathcal{U}^{(\pm)}_{(>)} (r)$.
This Green's function
technique has also been used for the study of the related strong-coupling inverse square potential 
regularization~\cite{Green_operator_approach}.

For Eq.~(\ref{eq:gen-generator_diff-eq}),
from the general solution~(\ref{eq:Whittaker-transformed-DE-HO_sol}), we identify 
\begin{equation} 
\mathcal{U}^{(\pm)}_{(<)} (r)
=  r^{-1/2} \,
 \mathcal{M}_{ \pm i \sigma \mathfrak{g} /\sqrt{\Delta},\mu/2} 
 ( \mp i  \check{r}^{2} ) 
\; 
\label{eq:Whittaker-transformed-DE-HO_SL<}
\end{equation}
as the solutions that satisfy a regular boundary condition $\left. \mathcal{U}(r) \right|_{r=0}  =0$ at the origin; and
\begin{equation}
\mathcal{U}^{(\pm)}_{(>)} (r)
=  r^{-1/2} \,
{W}_{ \pm i  \sigma \mathfrak{g} /\sqrt{\Delta},\mu/2} 
 ( \mp i  \check{r}^{2} ) 
\; 
\label{eq:Whittaker-transformed-DE-HO_SL>}
\end{equation}
as the solutions that satisfy the appropriate asymptotic boundary condition at infinity 
(outgoing/incoming waves for $G^{(\pm)}$).
The identity~(\ref{eq:Whittaker-M-analytic-cont}) shows that 
$\mathcal{U}^{(\pm)}_{(<)} \propto  \mathcal{M}_{ \pm i \sigma \mathfrak{g} /\sqrt{\Delta},\mu/2} $ are actually the same function, up to a proportionality constant; however, 
$\mathcal{U}^{(\pm)}_{(>)}  \propto W_{ \pm i  \sigma \mathfrak{g} /\sqrt{\Delta},\mu/2} $ are distinctly
different. The Wronskian can be computed with the identity
$\mathfrak{W} \left\{ 
 \mathcal{M}_{ \lambda,\mu/2}  ,
 {W}_{\lambda,\mu/2} 
 \right\}  (z) = 
- \left[  \Gamma \left( (1+\mu)/2 - \lambda \right) \right]^{-1} $; 
using Eqs.~(\ref{eq:Whittaker-transformed-DE-HO_SL<}) and (\ref{eq:Whittaker-transformed-DE-HO_SL>}),
along with  $p(r) =1$,
and applying the chain rule for $z=\mp i  \check{r}^{2}$, 
the final result for ${\mathcal G}^{(\pm)}_{l +\nu}(r'',r';E)$ is identical to
 Eq.~(\ref{eq:GF_RHO}). This again verifies the equivalence of the path-integral and differential-equation approaches,
 and provides additional consistency checks for the network of relations defined in this paper.

\end{document}